\documentclass[useAMS,usenatbib]{mn2e}

\usepackage{graphicx}
\usepackage{epstopdf}
\title[Resurrecting the Red from the Dead: Optical Properties of BCGs in X-ray Luminous Clusters]{Resurrecting the Red from the Dead: Optical Properties of BCGs in X-ray Luminous Clusters\thanks{Based on observations obtained at the Canada-France-Hawaii Telescope (CFHT) which is operated by the National Research Council of Canada, the Institut National des Sciences de l'Univers of the Centre National de la Recherche Scientifique of France, and the University of Hawaii.}}

\author[Bildfell et al.]{Chris Bildfell\thanks{E-mail: bildfell@uvic.ca (CB); hoekstra@uvic.ca (HH); babul@uvic.ca (AB); amahdavi@uvic.ca (AM)}, Henk Hoekstra\thanks{Alfred P. Sloan fellow}, Arif Babul, Andisheh Mahdavi\\
Department of Physics \& Astronomy, University of Victoria, Victoria, BC V8P 1A1, Canada}

\begin{document} 

\date{Submitted year month day}

\pagerange{\pageref{firstpage}--\pageref{lastpage}} \pubyear{2007}

\maketitle

\label{firstpage}

\begin{abstract}
We present measurements of surface brightness and colour profiles for the brightest cluster galaxies (BCGs) in a sample of 48 X-ray luminous galaxy clusters.  These data were obtained as part of the Canadian Cluster Comparison Project (CCCP).  The Kormendy relation of our BCGs is steeper than that of the local ellipticals, suggesting differences in the assembly history of these types of systems.  We also find that while most BCGs show monotonic colour gradients consistent with a decrease in metallicity with radius, 25\% of the BCGs show colour profiles that turn bluer towards the centre (blue-cores).   We interpret this bluing trend as evidence for recent star formation.  The excess blue light leads to a typical offset from the red sequence of $0.5$ to $1.0$ $mag$ in ($g'$-$r'$), thus affecting optical cluster studies that may reject the BCG based on colour.  All of the blue-core BCGs are located within $\sim10$kpc of the peak in the cluster X-ray emission.  Furthermore, virtually all of the BCGs with recent star formation are in clusters that lie above the $L_x-T_x$ relation.  Based on photometry alone, these findings suggest that central star formation is a ubiquitous feature of BCGs in dynamically relaxed cool-core clusters.  This implies that while AGNs and other heating mechanisms are effective at tempering cooling, they do not full compensate for the energy lost via radiation.
\end{abstract}


\section{Introduction}

\indent  It is well known that the hot intra-cluster medium (ICM) in galaxy clusters radiates enormous amounts of energy at X-ray wavelengths ($L_x \sim 10^{44} - 10^{46} erg s^{-1}$).  Taken at face value, this loss of energy from the ICM implies that the gas should be cooling, losing pressure support and flowing inward.  Early cooling flow models predicted cooling and star formation rates on the order of $300 \la \dot{M} \la 1000$ $M_{\odot}$ yr$^{-1}$ (Fabian 1994).  The lack of widespread detection of iron lines expected from cluster gas cooling below 1-2 keV in XMM-Newton observations of cool core clusters (Peterson et al. 2003) contradicted this model and created the need to incorporate non-gravitational heating mechanisms to counterbalance the radiative energy loss.  While these observations have effectively ruled out accretion flows large enough to form an $L_*$ galaxy in 100 Myr, they do not preclude that star formation is ongoing in cool core clusters at a much reduced rate.  The level of star formation still allowed is uncertain and it is important to investigate this through means independent of the X-ray-determined mass flow rate.  Of particular interest is the distinction between models in which heating completely balances cooling and those in which heating offsets most but not all of the radiative losses.  If significant star formation is occurring as a result of a heating imbalance it should be reflected in the optical properties of the brightest cluster galaxy (BCG), which is often found deep within the cluster potential.  A good way to test this is to look for links between the BCG optical properties with the X-ray properties of their host clusters.  In fact, several case studies over the last two decades have found evidence for recent star formation in the central galaxies of cool core clusters (Johnstone et al. 1987, McNamara \& O'connell 1989, 1992, Allen et al. 1992, Donahue \& Voit 1997, Cardiel et al. 1998, Crawford et al. 1999, Edge 2001, Edge et al. 2002, Goto 2005, McNamara et al 2006, Wilman et al. 2006, Edwards et al. 2007).  Furthermore, the strength of the recent star formation indicators is at least weakly correlated with their cooling-flow derived mass drop out rates (Johnstone et al. 1987, Heckman et al. 1989, McNamara 1997).

\begin{table*} 
\centering
\begin{minipage}{175mm}
\caption{CCCP Targets Observed with MegaCam}
\begin{tabular}{|l|l|l|l|l|l|l|l|l|c|}
\hline
 Name & $z$ & $\alpha_{BCG}$ & $\delta_{BCG}$ & $r_e^{r'}$ & $\mu_e^{r'}$ & $R_{\rmn{off}}$ & Core & $R_*$ & BCG \\
 & & (J2000) & (J2000) & kpc & mag/$(")^2$ & kpc & colour & kpc & Code \\
\hline
\hline
3C295 & 0.46 & 14 11 20.57 & $+52$ 12 09.93 & $31.91^{+.89}_{-.79}$ & $23.283^{+.055}_{-.051}$ & 12.0 & blue & $15\pm3$ & 1 \\
Abell 115 & 0.20 & 00 56 00.26 & $+26$ 20 32.65 & $41.09^{+.14}_{-.12}$ & $23.937^{+.005}_{-.005}$ & 143 & red & - & 1 \\
Abell 115N & 0.20 & 00 55 50.62 & $+26$ 24 37.55 & $14.94^{+.05}_{-.07}$ & $22.301^{+.005}_{-.006}$ & 10.1 & blue & $11\pm1$ & 1 \\
Abell 223a & 0.21 & 01 38 02.29 & $-12$ 45 19.92 & $42.04^{+.49}_{-.42}$ & $23.852^{+.022}_{-.019}$ & - & red & - & 1 \\
Abell 223b & 0.21 & 01 37 55.99 & $-12$ 49 10.09 & $76.66^{+.56}_{-.51}$ & $25.267^{+.010}_{-.010}$ & 8.20 & red & - & 1 \\
Abell 520 & 0.20 & 04 54 14.06 & $+02$ 57 10.63 & $40.36^{+.37}_{-.42}$ & $24.156^{+.016}_{-.019}$ & 340 & red & - & 2 \\
Abell 521 & 0.25 & 04 54 06.89 & $-10$ 13 24.61 & $37.11^{+.47}_{-.51}$ & $23.614^{+.021}_{-.025}$ & 32.8 & red & - & 1 \\
Abell 586 & 0.17 & 07 32 20.31 & $+31$ 38 01.06 & $57.96^{+.47}_{-.54}$ & $24.038^{+.013}_{-.013}$ & 10.7 & red & - & 1 \\
Abell 611 & 0.29 & 08 00 56.83 & $+36$ 03 23.79 & $34.40^{+.33}_{-.28}$ & $23.068^{+.017}_{-.015}$ & 4.0 & red & - & 1 \\
Abell 697 & 0.28 & 08 42 57.58 & $+36$ 21 59.38 & $70.54^{+.54}_{-.47}$ & $24.396^{+.012}_{-.010}$ & 20.3 & red & - & 1 \\
Abell 851 & 0.41 & 09 42 57.47 & $+46$ 58 49.94 & $17.27^{+.20}_{-.98}$ & $22.306^{+.046}_{-.171}$ & 278 & blue & $10\pm1$ & 3 \\
Abell 959 & 0.29 & 10 17 36.03 & $+59$ 34 01.81 & $10.67^{+0.05}_{-.09}$ & $22.269^{+.012}_{-.021}$ & 36.5 & red & - & 2 \\
Abell 1234 & 0.166 & 11 22 29.95 & $+21$ 24 22.00 & $21.47^{+.24}_{-.22}$ & $23.019^{+.022}_{-.020}$ & - & red & - & 2 \\
Abell 1246 & 0.19 & 11 23 58.83 & $+21$ 28 49.84 & $55.00^{+.37}_{-.33}$ & $24.490^{+.012}_{-.010}$ & 48.4 & red & - & 2 \\
Abell 1758b & 0.28 & 13 32 38.43 & $+50$ 33 36.01 & $51.53^{+.33}_{-.42}$ & $24.227^{+.011}_{-.014}$ & 25.2 & red & - & 1 \\
Abell 1835 & 0.25 & 14 01 02.10 & $+02$ 52 42.69 & $66.39^{+.56}_{-.51}$ & $24.125^{+.016}_{-.015}$ & 6.4 & blue & $19\pm2$ & 1 \\
Abell 1914 & 0.171 & 14 26 03.90 & $+37$ 49 53.57 & $49.16^{+.23}_{-.28}$ & $24.397^{+.008}_{-.009}$ & 85.6 & red & - & 2 \\
Abell 1942 & 0.22 & 14 38 21.88 & $+03$ 40 13.34 & $43.69^{+.22}_{-.22}$ & $23.549^{+.008}_{-.009}$ & 4.30 & red & - & 1 \\
Abell 2104 & 0.16 & 15 40 07.94 & $-03$ 18 16.25 & $56.48^{+.37}_{-.37}$ & $24.244^{+.012}_{-.013}$ & 7.66 & red & - & 1 \\
Abell 2111 & 0.23 & 15 39 40.52 & $+34$ 25 27.46 & $26.42^{+.35}_{-.31}$ & $23.206^{+.025}_{-.023}$ & 129 & red & - & 2 \\
Abell 2163 & 0.20 & 16 15 48.99 & $-06$ 08 41.21 & $26.33^{+.56}_{-.47}$ & $23.148^{+.036}_{-.031}$ & 160 & red & - & 1 \\
Abell 2204 & 0.15 & 16 32 46.98 & $+05$ 34 32.84 & $130.13^{+.49}_{-.33}$ & $25.006^{+.006}_{-.004}$ & 1.15 & blue & $14\pm3$ & 1 \\
Abell 2259 & 0.16 & 17 20 09.66 & $+27$ 40 08.29 & $44.94^{+.51}_{-.42}$ & $23.599^{+.021}_{-.017}$ & 78.2 & red & - & 1 \\
Abell 2261 & 0.22 & 17 22 27.23 & $+32$ 07 57.72 & $36.10^{+.14}_{-.14}$ & $22.897^{+.007}_{-.008}$ & 0.39 & red & - & 1 \\
Abell 2537 & 0.30 & 23 08 22.22 & $-02$ 11 31.74 & $46.37^{+.42}_{-.47}$ & $23.605^{+.016}_{-.018}$ & 17.0 & red & - & 1 \\
CL0910+41 & 0.44 & 09 13 45.52 & $+40$ 56 28.54 & $32.64^{+.49}_{-.42}$ & $23.358^{+.029}_{-.025}$ & 3.87 & blue & $40\pm8$ & 1 \\
CL1938+54 & 0.26 & 19 38 18.12 & $+54$ 09 40.19 & $148.42^{+1.7}_{-1.4}$ & $25.298^{+.019}_{-.016}$ & - & red & - & 1 \\
MS0440+02 & 0.19 & 04 43 09.92 & $+02$ 10 19.33 & $102.00^{+.33}_{-.28}$ & $24.458^{+.005}_{-.004}$ & 0.87 & red & - & 1 \\
MS0451-03 & 0.54 & 04 54 10.84 & $-03$ 00 51.39 &$45.55^{+.29}_{-.58}$ & $24.418^{+.015}_{-.024}$ & 28.4 & blue & $20\pm2$ & 1 \\
MS1008-12 & 0.30 & 10 10 32.34 & $-12$ 39 52.77 & $33.10^{+1.5}_{-1.4}$ & $23.214^{+.073}_{-.072}$ & 9.88 & red & - & 1 \\
RXJ1347.5-1145 & 0.45 & 13 47 30.65 & $-11$ 45 09.09 & $29.61^{+.16}_{-.16}$ & $23.154^{+.012}_{-.014}$ & 1.64 & blue & $20\pm2$ & 1 \\
\hline
\end{tabular}

{\small BCGs (by cluster) that were observed with MegaCam.  Spectroscopically determined redshifts ($z$) of the host clusters are shown.  The coordinates of the selected BCG are given as ($\alpha_{BCG},\delta_{BCG}$); see section \ref{selection} in the appendix for details.  The $r'$ best fit surface brightness profile parameters $r_e^{r'}$ and $\mu_e^{r'}$ are given.  The value of $\mu_e^{r'}$ has been corrected for galactic extinction and cosmological dimming.  The projected offset of the BCG with respect to the peak in host cluster X-ray emission is given as $R_{\rmn{off}}$.  Core colour is given as "red" or "blue" based on the shape of the inner colour profile (see section \ref{colourprodiscuss}).  If a blue core is present the size of the blue region is given as $R_*$.  Selection code refers to the degree of ambiguity assessed to the BCG selection; [1] obvious (cD), [2] non-trivial (X-ray necessary), [3] highly ambiguous.}
\label{tab:MegaCamdata}
\end{minipage}
\end{table*}
\normalsize

\begin{table*}

\centering
\begin{minipage}{175mm}
\caption{CCCP Targets Observed with CFH12K}
\begin{tabular}{|l|l|l|l|l|l|l|l|l|c|}
\hline
 Name & $z$ & $\alpha_{BCG}$ & $\delta_{BCG}$ & $r_e^R$ & $\mu_e^R$ & $R_{\rmn{off}}$ & Core & $R_*$ & BCG \\
 & & (J2000.0) & (J2000.0) & kpc & mag/$(")^2$ & kpc & colour & kpc & Code \\
\hline
\hline
Abell 68 & 0.255 & 00 37 06.85 & $+09$ 09 24.51 & $85.76^{+.79}_{-.84}$ & $24.780^{+.015}_{-.016}$ & 15.4 & blue & $16\pm3$ & 1 \\
Abell 209 & 0.206 & 01 31 52.54 & $-13$ 36 40.00 & $89.788^{+.56}_{-.51}$ & $24.394^{+.011}_{-.009}$ & 16.4 & red & - & 1 \\
Abell 267 & 0.23 & 01 52 41.95 & $+01$ 00 25.89 & $76.69^{+.70}_{-.70}$ & $23.899^{+.017}_{-.016}$ & 77.5 & red & - & 1 \\
Abell 370a & 0.375 & 02 39 52.70 & $-01$ 34 18.52 & $110.9^{+2.3}_{-2.3}$ & $24.633^{+.025}_{-.027}$ & 193 & blue & $20\pm2$ & 2 \\
Abell 370b & 0.375 & 02 39 53.10 & $-01$ 34 55.71 & $28.48^{+.62}_{-.62}$ & $22.812^{+.042}_{-.042}$ & 23.1 & red & - & 2 \\
Abell 383 & 0.187 & 02 48 03.38 & $-03$ 31 44.93 & $26.95^{+.16}_{-.14}$ & $22.502^{+.011}_{-.010}$ & 0.71 & blue & $28\pm5$ & 1 \\
Abell 963 & 0.206 & 10 17 03.63 & $+39$ 02 49.67 & $60.00^{+.35}_{-.35}$ & $23.679^{+.010}_{-.010}$ & 6.33 & red & - & 1 \\
Abell 1689 & 0.1832 & 13 11 29.52 & $-01$ 20 27.86 & $46.64^{+.35}_{-.27}$ & $23.349^{+.014}_{-.011}$ & 5.01 & red & - & 1 \\
Abell 1763 & 0.223 & 13 35 20.12 & $+41$ 00 04.30 & $24.47^{+.19}_{-.21}$ & $22.548^{+.014}_{-.015}$ & 7.40 & red & - & 1 \\
Abell 2218 & 0.1756 & 16 35 49.26 & $+66$ 12 44.56 & $46.34^{+.39}_{-.47}$ & $23.602^{+.014}_{-.017}$ & 59.5 & red & - & 1 \\
Abell 2219 & 0.2256 & 16 40 19.85 & $+46$ 42 41.30 & $61.66^{+.51}_{-.51}$ & $23.941^{+.014}_{-.014}$ & 8.27 & red & - & 1 \\
Abell 2390 & 0.2280 & 21 53 36.84 & $+17$ 41 44.10 & $18.42^{+.26}_{-.23}$ & $22.540^{+.029}_{-.026}$ & 3.63 & blue & $25\pm5$ & 1 \\
CL0024+16 & 0.39 & 00 26 35.68 & $+17$ 09 43.48 & $60.1^{+1.8}_{-1.8}$ & $23.836^{+.042}_{-.044}$ & 24.5 & red & - & 3 \\
MS0016+16 & 0.5465 & 00 18 33.56 & $+16$ 26 16.30 & $42.38^{+1.1}_{-1.2}$ & $23.797^{+.044}_{-.048}$ & 41.2 & - & - & 3 \\
MS0906+11 & 0.1704 & 09 09 12.76 & $+10$ 58 29.12 & $55.35^{+.37}_{-.33}$ & $23.809^{+.012}_{-.011}$ & 2.55 & red & - & 2 \\
MS1224+20 & 0.3255 & 12 27 13.48 & $+19$ 50 56.10 & $36.57^{+.70}_{-.70}$ & $23.379^{+.036}_{-.038}$ & - & red & - & 1 \\
MS1231+15a & 0.2353 & 12 33 54.72 & $+15$ 26 19.96 & $51.93^{+.51}_{-.61}$ & $24.359^{+.015}_{-.019}$ & - & red & - & 3 \\
MS1231+15b & 0.2353 & 12 33 55.35 & $+15$ 25 59.11 & $50.97^{+.35}_{-.42}$ & $24.100^{+.012}_{-.014}$ & - & red & - & 3 \\
MS1358+62 & 0.3290 & 13 59 50.62 & $+62$ 31 05.09 & $48.32^{+.79}_{-.88}$ & $24.071^{+.029}_{-.032}$ & 4.41 & red & - & 1 \\
MS1455+22 & 0.2568 & 14 57 15.12 & $+22$ 20 34.48 & $24.80^{+.47}_{-.53}$ & $22.220^{+.033}_{-.038}$ & 3.29 & blue & $29\pm3$ & 1 \\
MS1512+36 & 0.3727 & 15 14 22.51 & $+36$ 36 21.30 & $57.56^{+.86}_{-.86}$ & $24.111^{+.024}_{-.024}$ & 5.74 & blue & $15\pm2$ & 1 \\
MS1621+26a & 0.4275 & 16 23 35.17 & $+26$ 34 28.40 & $116.1^{+3.5}_{-3.7}$ & $26.004^{+.036}_{-.042}$ & 74.7 & - & - & 2 \\
MS1621+26b & 0.4275 & 16 23 35.45 & $+26$ 34 14.50 & $46.5^{+1.1}_{-1.2}$ & $24.193^{+.038}_{-.047}$ & 41.2 & blue & $21\pm5$ & 2 \\
\hline
\end{tabular}

{\small BCGs (by cluster) that were observed with CFH12K.  Spectroscopically determined redshifts ($z$) of the host clusters are shown.  The coordinates of the selected BCG are given as ($\alpha_{BCG},\delta_{BCG}$); see section \ref{selection} in the appendix for details.  The $r'$ best fit surface brightness profile parameters $r_e^{R}$ and $\mu_e^{R}$ are given.  The value of $\mu_e^{R}$ has been corrected for galactic extinction and cosmological dimming.  The projected offset of the BCG with respect to the peak in host cluster X-ray emission is given as $R_{\rmn{off}}$.  Core colour is given as "red" or "blue" based on the shape of the inner colour profile (see section \ref{colourprodiscuss}).  If a blue core is present the size of the blue region is given as $R_*$.  Selection code refers to the degree of ambiguity assessed to the BCG selection; [1] obvious (cD), [2] non-trivial (X-ray necessary), [3] highly ambiguous.}
\label{tab:CFH12Kdata}
\end{minipage}
\end{table*}
\normalsize

\indent  However, despite the compelling evidence for recent star formation in the BCGs of cool core clusters, the source of the cold gas required to fuel this star formation is yet uncertain.  Two competing explanations are that it is a product of radiative cooling of the ICM (ie. cooling flow) or that the cold gas is deposited during a merger event (\textit{c.f.} Poole et al. 2006).  The uncertainty surrounding the origin of the cold gas persists mainly because previous work focused on samples of cool core clusters.  However, to resolve this issue requires a comparison of BCGs across a wide variety of host cluster X-ray morphologies (ie. cool cores and non-cool cores).  We therefore undertake a comprehensive study of BCGs using a cluster sample that spans the full observed range of scatter in the cluster X-ray scaling relations (ie. mass and X-ray luminosity vs. X-ray temperature).

\indent We present here an analysis of the surface brightness and colour profiles for 48 BCGs.  We compare these optical data to the global X-ray luminosity ($L_x$) and temperature ($T_x$) of their host clusters.  We aim to discover how the recent star formation in BCGs is linked to the balance between heating and cooling clusters.  This information is crucial to our understanding of how baryonic feedback influences the evolution of the ICM.

\indent The structure of the paper is as follows.  We describe the data in section \ref{data}.  Section \ref{sbprodiscuss} contains a discussion of the BCG surface brightness profiles.  In section \ref{colourprodiscuss} we discuss BCG colour profiles.  Section \ref{xray} gives a comparison of BCG optical properties to global cluster X-ray properties.  The conclusions are contained in section \ref{conclusion}.  The cosmological parameters used throughout this work are $H_0=70$ km s$^{-1}$ Mpc$^{-1}$, $\Omega_m=0.3$ and $\Omega_{\Lambda}=0.7$.

\section{Data} \label{data}

\indent The data studied here are part of the Canadian Cluster Comparison Project (CCCP)\footnote{Based on observations obtained with MegaPrime/MegaCam, a joint project of CFHT and CEA/DAPNIA, at the Canada-France-Hawaii Telescope (CFHT) which is operated by the National Research Council (NRC) of Canada, the Institute National des Sciences de l'Univers of the Centre National de la Recherche Scientifique of France, and the University of Hawaii.}.  The CCCP is a study of an X-ray selected sample of 50 clusters from the ASCA catalog of Horner (2001) with redshifts in the range $0.15 \leq z \leq 0.55$ and most having X-ray temperatures $T_x > 5$ keV.  The CCCP is selected such that it spans the full range of observed scatter in the cluster X-ray and Sunyaev-Zel'dovich (SZ) scaling relations.  This was done to create maximal leverage for the investigation of baryonic feedback effects.  The combination of X-ray and deep, multi-filter optical data makes the CCCP an excellent sample for studying the links between BCGs and their host cluster environments.  The deep optical imaging data was obtained using the Canada-France-Hawaii Telescope (CFHT).  The full sample comprises a set of 30 clusters observed in $g'$ and $r'$ using the MegaCam detector and 20 clusters observed in the $B$ and $R$ filters using the CFH12K detector which were found in the CFHT data archive at the Canadian Astronomical Data Centre (CADC) (see Hoekstra (2007) for more details).  MegaCam and CFH12K have 0.186$''/$pix and 0.206$''/$pix image scales respectively and with the median seeing of 0.7$''$ at the CFHT prime focus these cameras image the central $r\la 35 h^{-1}_{70}$ kpc regions of the target BCGs ($\sim 10''$ at a typical redshift of the clusters in our sample) over several resolution elements (see figure \ref{fig:sb31_mega}).  All of the data are taken with between $0.5"$ and $1.0"$ seeing.  The FWHM of the PSFs for the individual images are shown on the surface brightness and colour profiles in appendix \ref{photo}.

\indent  Table \ref{tab:MegaCamdata} lists all targets observed with MegaCam and Table \ref{tab:CFH12Kdata} lists the targets observed with CFH12k.  For the clusters imaged with MegaCam, the ones with z $<$ 0.4 were observed for 4 $\times$ 450 seconds in $g'$ and 8 $\times$ 600 seconds in $r'$ while clusters with z $>$ 0.4 were observed for 4 $\times$ 600 seconds in $g'$ and 8 $\times$ 600 seconds in $r'$.  The clusters that were imaged with CFH12K were taken with a variety of exposure times, ranging from 3000 to 15300 seconds in $B$  and from 3000 to 13800 seconds in R.  The exact exposure times for the individual CFH12K clusters can be found in Hoekstra (2007).
  
\indent  The initial data reduction (bias subtraction and flat-fielding) are done on site at CFHT with the Elixir automated instrument and software pipeline\footnote{\texttt{http://www.cfht.hawaii.edu/Instruments/Elixir/}}.  The pipeline also provides photometric zero-points based on observations of photometric standard stars during the observing run.  The images of each cluster are aligned and median-stacked using the SWarp software package.

\indent  The BCG selection is carried out using the combined optical and X-ray data sets (see Appendix \ref{selection} for details).

\section{Surface Brightness Profiles} \label{sbprodiscuss}

\indent  In this section we discuss the surface brightness profiles of BCGs in the 48 CCCP clusters.  We measure the surface brightness in circular annuli.  The profiles are presented in appendix \ref{photo}.

\indent  We fit circularly symmetric, PSF-convolved $r^{1/4}$ models (deVaucouleurs 1948) to the surface brightness profiles of the CCCP BCGs, which yield estimates for the effective radius ($r_e$) and effective surface brightness ($\mu_e$).  Details of the fitting procedure are given in appendix \ref{sbfit}.  Individual fits to each system are shown in figures \ref{fig:sb31_mega} and \ref{fig:sb23_12k}.  The model fits are of varying quality among the CCCP BCGs.  Most BCGs are well fit, but exhibit deviations from a pure deVaucouleurs profile at either small or large radii.  We note that more complicated models (e.g. S\`ersic $r^{1/n}$ or double-$r^{1/4}$) may fit the profiles better but for our purposes the improvement is negligible.  For further discussion on the validity of various choices of surface brightness models see Gonzalez et al. (2005) and references therein.

\indent  Some of the observed profiles exhibit steep changes in the surface brightness over a small range in radius (eg. Abell 223a, Abell 521, Abell 697, CL1938+54).  These steep changes in brightness are caused by imposing a circularly symmetric profile on an object which is intrinsically elliptical.  These sharp variations occur wherever large portions of the BCG major or minor axis are masked.  For instance, if the major axis is heavily masked in a given radial bin, the mean surface brightness in that bin is driven towards a value that is more representative of surface brightness along the minor axis.  The impact of this effect is most noticeable for those objects which are highly elliptical (CL1938+54) and also in the case where opposing sides along the major/minor axis are heavily masked in the same radial bin.

\subsection{Kormendy Relation} \label{kormendy}

\indent The Kormendy Relation (KR) (Kormendy 1977) is an empirical scaling relation between $\mu_e$ and $r_e$ for elliptical galaxies and has been shown to be a projection of the more general Fundamental Plane (PF) (Djorgovski and Davis 1987, Dressler et al. 1987),  which describes a 2-dimensional manifold in ($\sigma_0$,$\langle\mu\rangle_e$,$r_e$) space.  The KR was first observed to hold for local ellipticals and has since been applied to a large range of spheroid galaxy types.

\begin{figure}
   \centering
   \includegraphics[width=3.5in]{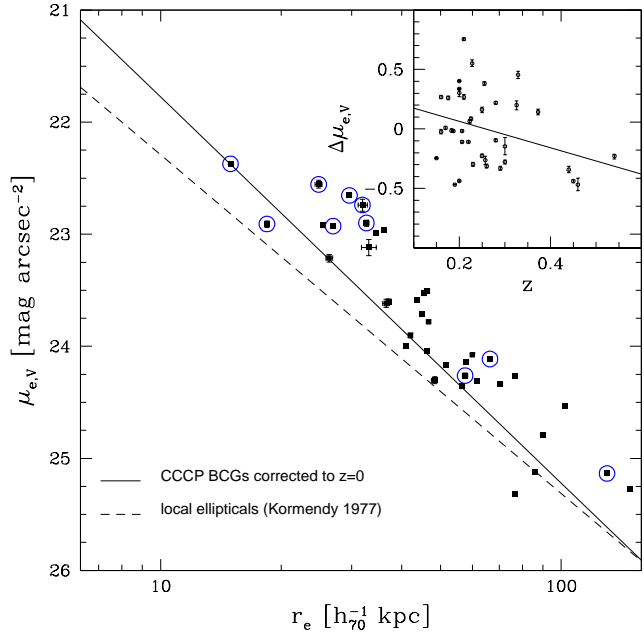} 
   \caption[Kormendy Relation for the CCCP BCGs]{Kormendy Relation (KR) for the CCCP BCGs.  The data has been corrected for cosmological dimming and converted to rest-frame $V$ magnitude.  Black squares denote the best fit $r^{1/4}$ model parameters.  Error bars show the 1 $\sigma$ fitting uncertainties but are typically smaller than the symbol size.  The solid black line shows the mean relation for the CCCP BCGs at $z=0$ after correcting for passive surface brightness evolution.  The dashed, black line shows the extrapolated KR for local elllipticals (Kormendy 1977).  The circled points show the BCGs that host blue-cores (see section \ref{colourprodiscuss}).  The upper right panel shows the residual in effective surface brightness as a function of redshift.}
   \label{fig:mue_vs_re}
\end{figure}

\indent The KR for the CCCP BCGs is shown in figure \ref{fig:mue_vs_re}.  Surface brightness $\mu_e$ is corrected for the effects of cosmological dimming and galactic extinction.  Transformations to Johnson $V$ magnitudes are computed by applying an elliptical galaxy template spectrum (Coleman et al. 1980).  Only those parameters from the fits to the $r'$ and $R$ data are used as their $V$ magnitude conversions are less affected by the assumption of a particular spectral template compared to the $g'$ and $B$ data.  Only systems which are unambiguously selected as the BCG are shown (i.e. selection code=1, see section \ref{selection}).  The CCCP BCGs are plotted in black squares.  The residual in $\mu_e$ from the mean KR ($\Delta\mu_{e,V}$) versus redshift is shown in the inset of figure \ref{fig:mue_vs_re}.  If we assume that the offset is due to passive evolution we find a mean surface brightness evolution of:

\begin{equation}
\Delta\mu_{e,V} = \big(-1.11 \pm 0.40\big) z + \big(0.28 \pm 0.20\big)
\label{eqn:mue_evolution}
\end{equation}

\noindent The relation \ref{eqn:mue_evolution} is shown in the inset as a solid line and describes a slow dimming of the dominant stellar population as the BCG ages.  This type of passive evolutionary behavior is expected for massive galaxies that form the majority of their stellar content at relatively early cosmic times.  If we further make the assumption that BCGs are homologous this surface brightness evolution equates to a change in the mass-to-light ratio $M/L$ with time:

\begin{equation}
\Delta log M/L_V = -0.44 \pm0.16 \Delta z
\label{eqn:M/L_evolution}
\end{equation}

\noindent This result is in agreement with work by van Dokkum et al. (1998) who use FP observations of cluster ellipticals to obtain a $M/L$ evolution of $\Delta log M/L_B \cong -0.55 \pm 0.05 \Delta z$ for a $\Lambda$CDM cosmology.

\indent We apply the evolution correction (equation \ref{eqn:mue_evolution}) to the data and compute the $z=0$ KR for BCGs (thin solid black line).  The scatter in the evolution-corrected data $\sigma_{BCG}$ is 0.28, which is consistent with the scatter in the KR of local ellipticals (Kormendy (1977)).  The best fit slope of the KR for CCCP BCGs is $a_{BCG}=3.44\pm0.13$, which is steeper than the values of $a_{BCG}=3.125\pm0.14$ reported by Oegerle \& Hoessel (1991) and $a_E=3.02\pm0.14$ found for local ellipticals.  This is qualitatively consistent with the result of von der Linden et al. (2007) who argue that the difference in the KR slope between BCGs and regular ellipticals may be explained by a disparity in M/L or a break from homology between these two types of galaxies.  Such a difference in M/L indicates an increased dark matter fraction in BCGs, while a break from homology implies a difference in the way these two types of systems are assembled.

\indent  The points which are circled in figure \ref{fig:mue_vs_re} correspond to the BCGs that host blue cores (see section \ref{colourprodiscuss}).  Most of these systems have $r_e<35$ $h^{-1}_{70}$ kpc and in the CCCP sample they dominate the small $r_e$ end of the KR.  The majority of the blue-core systems lie on the bright side of the best fit to the data, but it is important to note that they do not lie significantly off the relation.  This implies that despite the blue emission in these systems, the surface brightness fit is more sensitive to the spatially extended, older stellar population.  We find that removing the blue-core BCGs from the KR leads to a further steepening of the slope to $a_{BCG}=3.66\pm0.15$.  The slope of the KR for the CCCP BCGs is thus in disagreement with that of local ellipticals, regardless of whether or not the blue-core systems are included in the analysis.

\section{colour Profiles} \label{colourprodiscuss}

\indent  Using our multi-filter photometry we construct colour profiles of the BCGs in our sample.  colour profiles trace the distribution of stellar populations in these systems, which is important for constraining formation and evolution models.  As shown in figures \ref{fig:c31_mega} and \ref{fig:c22_12k} in the appendix, the general trend seen in BCG colours is a gradual decline towards the blue with increasing radius.  This is expected for an age/metallicity gradient in massive ellipticals (Vader et al. 1988).

\indent  There are several exceptions, however, as 25\% ($13/53$) of BCGs in the CCCP show a reversal from this basic behavior with the colour becoming increasingly blue in the core.  These galaxies are displaced from their host-cluster red sequence by $\sim 0.5$ to $1.0$ $mag$ in ($g'$-$r'$), which is important for optical cluster studies that may reject the BCG based on colour.  The blue cores in these systems cannot be fully explained by AGN point-sources because they are at least twice the size of the seeing disk.  Two of the more extreme examples of blue-core BCGs are those in Abell 1835 and Abell 2204.  These systems are hosts of CO emission in their central regions (Edge 2001) and are found to host significant recent star formation (Crawford et al 1999, Wilman et al. 2006).  Four other BCGs in the MegaCam sample that show similar trends in their colour profiles are: the BCG in RXJ1347 (ie., the brightest known cluster in X-ray (Schindler et al. 1995)), the BCG in 3C295, which shows visible signs of strong AGN activity in X-ray and radio as well as signs of massive amounts of star formation (Thimm et al. 1994, Harris et al. 2000), the BCG in CL0910 which is identified as a hyper-luminous infrared galaxy (HLIRG; $L_{ir} > 10^{13}L_\odot$ Goto 2005), and the BCG in A115N which is a confirmed host of recent star formation (Crawford et al. 1999).  Of the BCGs in the CFH12K sample Abell 68, Abell 370a, Abell 383, Abell 2390, MS1455 and MS1512 are classified here as blue-core BCGs.

\indent  Table \ref{tab:SFRs} lists star formation rates derived from a literature search for a subset of blue-core BCGs.  The SFRs range from $\sim 1$ to $\sim 100$ $M_\odot$ yr$^{-1}$.  Although the table indicates that the typical SFR of a blue-core BCG is significant, it is important to note that it is an order of magnitude less than the SFR predicted by the classical cooling flow model.

\begin{figure}
   \centering
   \includegraphics[width=3.5in]{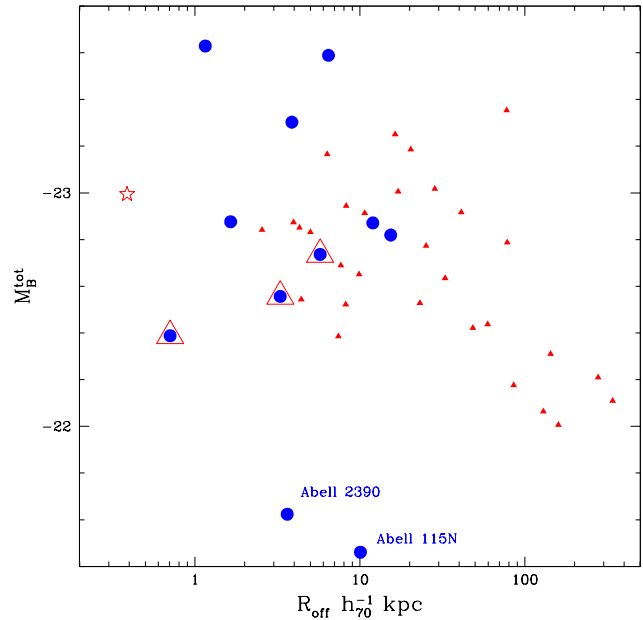} 
   \caption[BCG $M_B^{tot}$ vs. BCG/X-ray offset]{BCG $M_B^{tot}$ vs. BCG/X-ray offset.  Blue-core BCGs are identified with large blue circles.  The BCGs with host cluster $T_x < 5$ keV are shown with open red triangles.  There is an obvious tendency for brightest BCGs to lie closest to their host cluster's X-ray peak while the faintest BCGs are the furthest.  The starred symbol represents the BCG in Abell 2261 and is believed to host dust enshrouded star formation.  The anomalously faint blue-core BCGs belong to Abell 2390 and Abell 115N (see text for discusssion).}
   \label{fig:bcgmag_roffvsmt}
\end{figure}

\begin{table} 
\centering
\begin{minipage}{80mm}
\caption{Star formation rates derived from a literature search for a subset of blue-core BCGs}
\begin{tabular}{|p{45pt}|p{38pt}|p{55pt}|p{10pt}|}
\hline
 Name & $\dot{M}_*$ & Technique & Ref. \\
 & $M_\odot$ yr$^{-1}$ &  & \\
\hline
\hline
3C295 & 5.0 & [OII] \& [OIII] & (1) \\
Abell 115N & 0.2 & SSP (spectra) & (2) \\
                  & 6.2 & IR & (3) \\
Abell 1835 & 125 & IR  & (4) \\
                 &  40 &  H$\alpha$ & (4) \\
                 & 77 & SSP (spectra) & (2) \\
                 & 140$\pm$40 & SSP (U-R) & (6) \\
                 & 123$\pm$5 & UV excess & (7) \\
Abell 2204 & 14.7 & IR & (3) \\
                 & 1.3 & SSP (spectra) & (2) \\
Abell 2390 & 5 & IR & (4) \\
                  & 5 & H$\alpha$ & (4) \\
                  & 5.4 & SSP (spectra) & (2) \\
CL0910+41 &  41$\pm$12 & SDSS [0II]ew & (5) \\
MS1455+22 & 8.55  & SSP (spectra) & (2) \\
                    & 36$\pm4$ & UV excess & (7) \\

\hline
\end{tabular}

{\small  (1) Thimm et al. 1994, A\&A 285, 785 \\
(2) Crawford et al. 1999 MNRAS, 306, 857 \\
(3) Quillen et al. 2007, arXiv:0711.1118v1 \\
(4) Egami et al. 2006, ApJ 647, 922 \\
(5) SDSS DR6\\
(6) McNamara et al. 2006, ApJ, 648, 164\\
(7) Hicks \& Mushotzky 2005, ApJ, 635L, 9\\
\\
 Star formation rates (SFR) for several blue-core BCGs in the CCCP sample.  Technique refers to the method used to calculate the SFR.  For references (3) \& (5) the SFR was calculated using the indicated measure and the appropriate Kennicutt SFR conversion.  The rates shown here are significant, ranging from $\sim 1$ to $\sim 100$ $M_\odot$ yr$^{-1}$.  It is important to note, however, that the SFRs assembled here are an order of magnitude smaller than the estimates of the classical cooling flow model, which expected SFR ranging from  $\sim 100$ to $\sim 1000$ $M_\odot$ yr$^{-1}$

}
\label{tab:SFRs}
\end{minipage}
\end{table}
\normalsize

\subsection{Location within the Cluster} \label{location}

\indent  If the central star formation in the blue-core BCGs arises from gas that is cooling out of the ICM, then we expect these systems to be located exclusively at the centres of relaxed clusters, where the cool material is deposited.  To investigate the influence of cluster-centric distance on the luminosity of BCGs and the presence of blue-cores, we examine relationship between BCG total magnitude and the projected distance of the BCG from the cluster X-ray peak.

\indent  The X-ray peak positions are obtained using archival images from Chandra and XMM-Newton.  The total $B$-band magnitude of the BCG ($M_B^{tot}$) is calculated by integrating the measured $r'$ or $R$ surface brightness profile out to $r_e$ and doubling the resulting flux to account for the light outside $r_e$.  This total red flux is then converted to rest-frame $B$ magnitude, assuming an elliptical galaxy template spectrum of Coleman et al. (1980).

\indent Figure \ref{fig:bcgmag_roffvsmt} shows the BCG total B-band magnitude ($M_B^{tot}$) as a function of the projected offset from the host cluster X-ray peak ($R_{\rmn{off}}$).  The BCGs with blue cores are identified with large, blue circles.  It is immediately clear from this figure that all blue-core BCGs have small $R_{\rmn{off}}$; 90\% (9/10) of the blue-core BCGs have $R_{\rmn{off}}<10$ $h^{-1}_{70}$ kpc.  This is highly suggestive that the recent star formation in these systems is related to core processes within the cluster.  The system marked by a red star in this figure represents the BCG in Abell 2261.  From analysis of sub-mm images taken with SCUBA on JCMT, Chapman et al. (2002) find a significant amount of dust ($9.2 \times10^7 h_{70}^{-2} M_\odot $) associated with the BCG in A2261.  Our high-resolution imaging of this object reveals a double nucleus feature, possibly created by the presence of a dust lane.  These observations suggest that although this system lacks a blue core, it is either a host of dust-enshrouded star formation or that the conditions are ideal for the ignition of a starburst in this galaxy (ie. a large reservoir of cold gas).

\begin{figure*}
   \centering
   \begin{tabular}{c c}
   \includegraphics[width=3.5in]{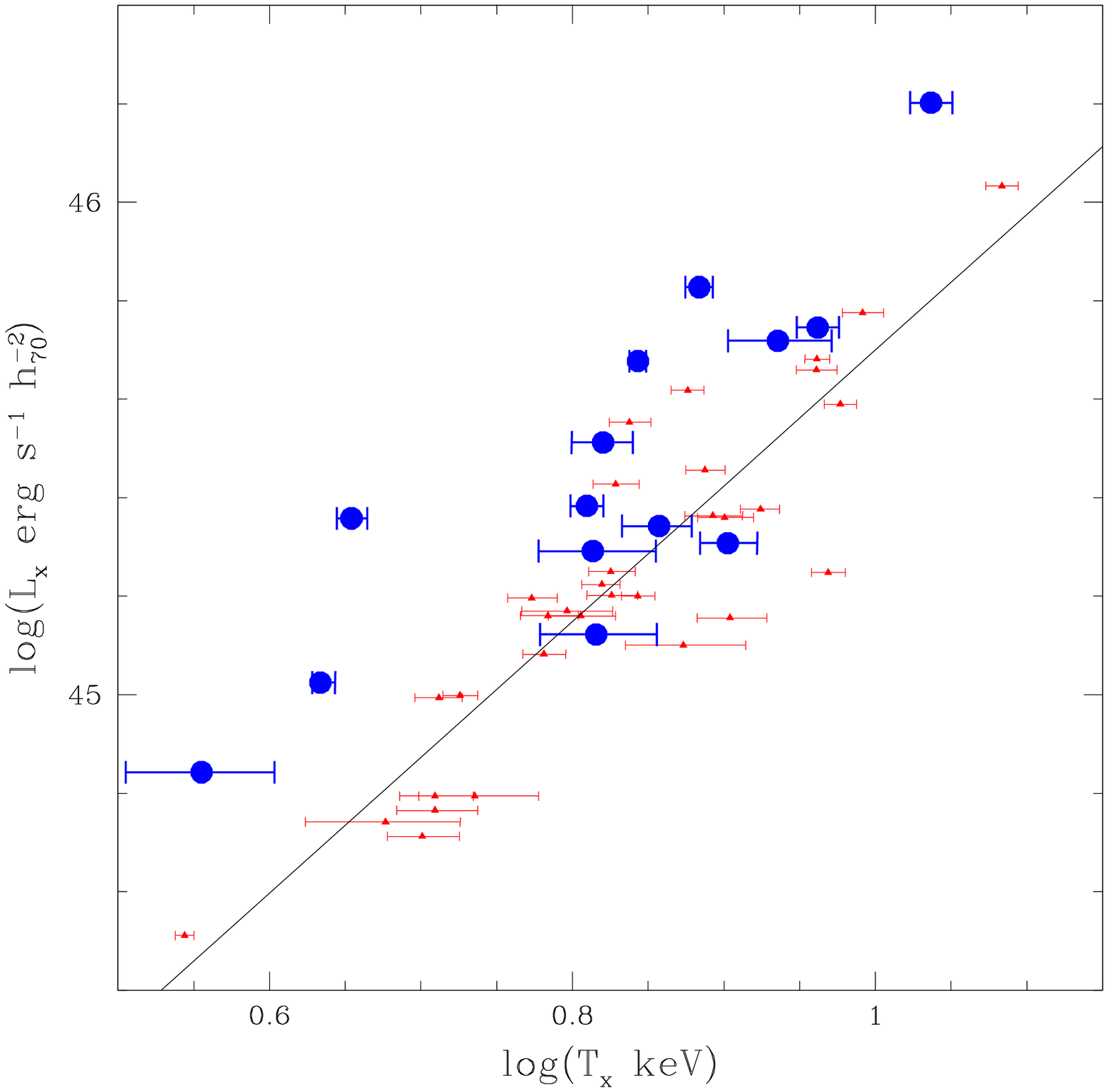} & \includegraphics[width=3.5in]{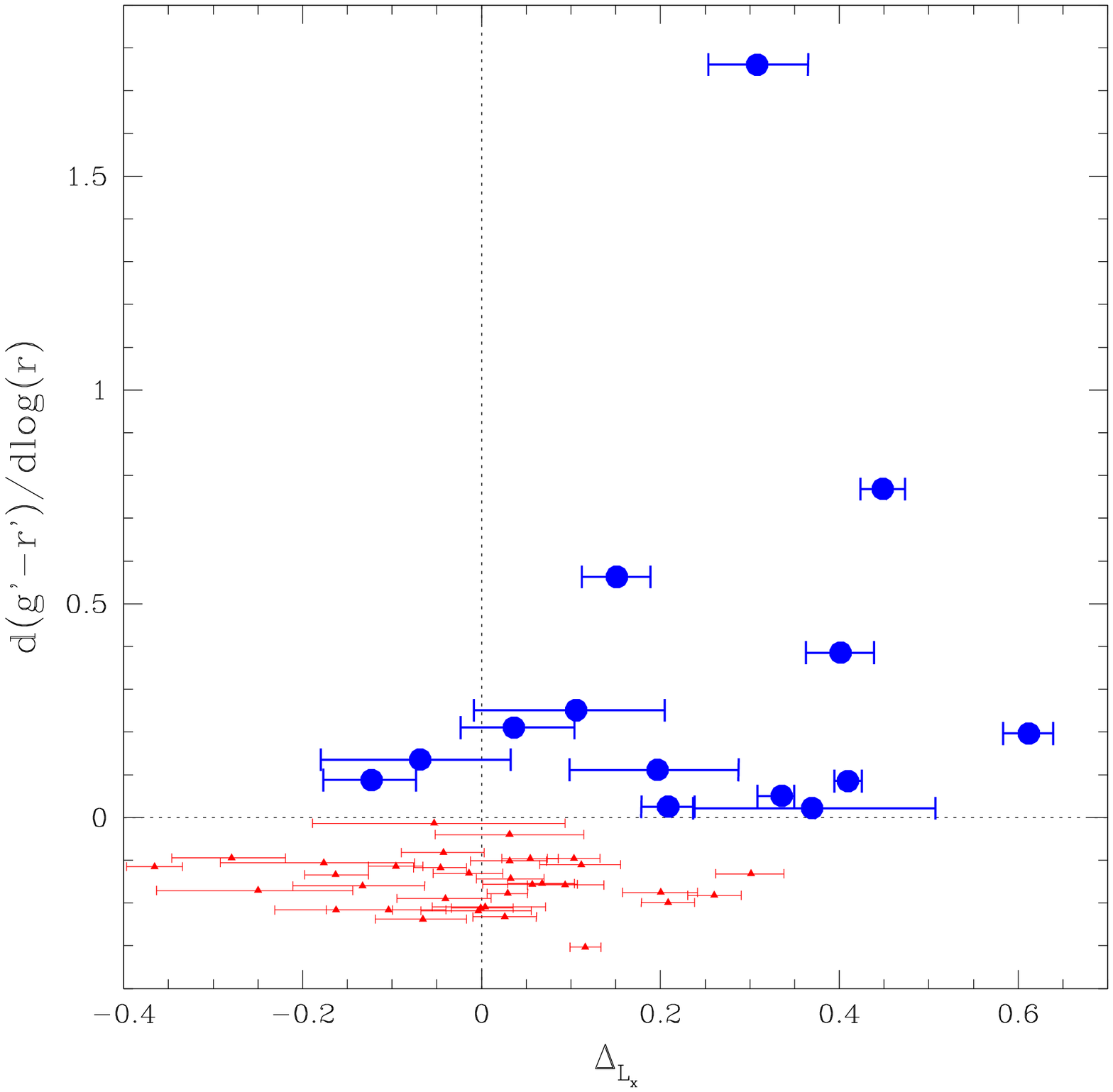}
   \end{tabular}
   \caption[The $L_x$-$T_x$ relation and BCG inner colour gradient for the CCCP clusters]{The $L_x$-$T_x$ relation and BCG inner colour gradient for CCCP clusters.  A simple measure of the inner colour gradient is obtained by performing a linear fit to $d(g'-r')/dlog(r)$ (MegaCam) or $d(B-R)/dlog(r)$ (CFH12K) between at $r=20$ $h^{-1}_{70}$ kpc and the FWHM of the limiting PSF.  Colour gradients in ($B-R$) are converted to ($g'-r'$) using the transformations of Fukugita et al. (1996).  Left: $L_x$-$T_x$ relation for the CCCP cluster.  Blue circles indicate cluster that contain BCGs with inverted central colour gradient (blue cores).  Right: colour gradient vs. offset in $L_x$ from the mean $L_x$-$T_x$ relation at a given $T_x$: $\Delta_{L_x}\equiv\Delta log(L_x/E(z)$ $ h^{-2}_{70} erg s^{-1})$ (X-ray excess).  Note the blue core BCGs are clustered at positive values of X-ray excess.}
   \label{fig:LTcolour}
\end{figure*}

\indent  The three systems marked with large, open, red triangles are the clusters with $T_x < 5$keV.  Because cluster mass and X-ray temperature are correlated, we expect these systems to be the least massive clusters in our sample.  Furthermore, since BCG luminosity scales with cluster mass (Lin \& Mohr 2004) these three BCGs are expected to be under-luminous compared to their counterparts in more massive clusters.  Despite the dynamic mass range of the CCCP sample, figure \ref{fig:bcgmag_roffvsmt} shows a general trend of increasing $M_B^{tot}$ with decreasing $R_{\rmn{off}}$.  There is a notable lack of bright BCGs with large $R_{\rmn{off}}$ as well as faint BCGs with small $R_{\rmn{off}}$.  This may be understood considering the following: the dependence of dynamical friction on satellite mass causes the more massive BCGs to sink to the centres of their clusters on shorter time scales than the less massive BCGs.  Furthermore, BCGs that are located in cluster centres grow more rapidly than those in the outer regions of cluster because they have an enhanced probability of trapping and cannibalizing smaller galaxies and in cool core clusters they are the beneficiaries of the cooling flows.  The combination of these effects will give rise to the trend in figure \ref{fig:bcgmag_roffvsmt}.

\indent  The obvious outliers from the trend in figure \ref{fig:bcgmag_roffvsmt} are the blue-core BCGs in Abell 2390 at $M_B^{tot} \approx -21.7$ and Abell 115N at $M_B^{tot} \approx -21.5$.  These systems have best-fit $r_e$ values of $18$ and $15$ kpc respectively; quite small compared to the rest of the BCG sample.  Neither of these galaxies have a well developed stellar envelope which could be the reason they are well fit with such compact surface brightness models.  The optical images of Abell 2390 show a wealth of substructure surrounding the BCG.  Bardeau et al. (2007) find that the weak lensing signal in this cluster requires an elongated mass mass distribution that matches the elongation seen in high-resolution X-ray imaging.  These results support the notion that the BCG in this cluster is young and has yet to cannibalize enough substructure to form a large cD envelope.  Abell 115N is the northern BCG in a binary cluster that appears to be in the early stages of a merger.  Metevier et al. (2000) find that this cluster has an anomalously high blue fraction for its redshift.  It may be that the reason the BCG in Abell 115N lacks a cD envelope is related to the age of the remainder of the galaxy population, which may not yet have had time to dynamically relax and become cannibalized.

\indent  We perform a Kendall's $\tau$ test on these data, which reveals that $M_B^{tot}$ and $R_{\rmn{off}}$ are correlated with a $\tau=0.26$ at the 98\% confidence level ($\sim 2.5 \sigma$ significance).  The interpretation of the trend seen in figure \ref{fig:bcgmag_roffvsmt} is complicated by the possible presence of selection effects, especially at the faint BCG/large offset end of the diagram.  Some of the clusters in the CCCP sample are in the early stages of a merger between two or more sub-clusters.  These systems naturally have large values of $R_{\rmn{off}}$ and do not have well developed cD galaxies.  Future work involving mass modeling of the clusters from the combined X-ray, optical, S-Z and weak lensing measurements will allow a more detailed investigation into these effects.  It is clear from examination of the X-ray profiles of the largest $R_{\rmn{off}}$ systems however, that not all of these objects are unrelaxed merging sub-clusters and thus the relation is not completely due to this type of selection bias

\section{Relation to X-ray properties of the host clusters} \label{xray}

\indent In this section we compare the optical properties of the BCGs in our sample with the X-ray properties of their host clusters.  The systems in the CCCP sample have been observed with ASCA and their X-ray temperatures and luminosities are taken from Horner (2001).

\indent  There is increasing evidence that the intrinsic scatter in the $L_x$-$T_x$ relation is physical in origin and that the processes that give rise to the scatter (ie. radiative cooling, star formation/stellar winds, AGN) are also responsible for the deviation in $M$-$L_x$ as well as SZ correlations (Babul et al. 2002, McCarthy et al. 2003, McCarthy et al. 2004, Balogh et al. 2006).

\indent McCarthy et al. (2004) demonstrated that models of cluster evolution that simultaneously incorporate radiative cooling along with a source of preheating reproduce a significant amount of the observed scatter in the cluster $L_x$-$T_x$ relation.  In these models, preheating drives a cluster of a given mass towards lower X-ray luminosities.  This creates an entropy floor in the cluster core which limits the central density and therefore $L_x$.  Radiative cooling can erase the entropy floor and produce dense, bright $L_x$ cluster cores.  These cool core clusters are expected to lie preferentially on the high luminosity side of the $L_x$-$T_x$ relation.

\indent  An interesting question is whether the same processes that are believed to cause the scatter in the $L_x$-$T_x$ relation also alter the optical properties of BCGs.  The CCCP is an excellent sample to investigate this with because it spans the full observed range of scatter in the cluster $L_x$-$T_x$ and $M$-$T_x$ relations.  To test this scenario we compare the deviation of cluster $L_x$ from the best fit $L_x$-$T_x$ to various properties of BCG.  The quantity $\Delta_{L_x}\equiv\Delta log(L_x/E(z)$ $ h^{-2}_{70} erg s^{-1})$, with $E(z)\equiv(\Omega_M(1+z)^3+\Omega_\Lambda)^{1/2}$, represents this deviation corrected for the change in cosmology and the evolution of the mean background density.  In what follows this metric is used as a first-order proxy for the cooling state of a cluster.

\subsection{Blue-core BCGs and the $L_x$-$T_x$ relation \label{Xraycolourgrad}}

\indent  In section \ref{location} we found that the BCGs hosting recent star formation are all located near the peak of their host cluster X-ray distribution.  We now seek to discover the differences, if any, between clusters that host blue-core BCGs and those that do not.  With this aim, we examine the inner colour gradients of the BCGs in the CCCP in relation to the cooling state proxy $\Delta_{L_x}$ described in section \ref{xray}.

\indent  The $L_x$-$T_x$ relations for the CCCP clusters is shown in the left hand panel of figure \ref{fig:LTcolour}.  The data have been corrected for any differences in cosmology and scaled by $E(z)$ to compensate for the evolution in the cosmic background density.  The mean relation, shown as a dotted line, is determined by fitting a power law of the form $L_x/E(z)=\beta T_x^\alpha$ to the clusters with regular BCGs (ie. red cores).  The large blue circles show those clusters that contain BCGs with positive central colour gradients (blue cores); where colour gradient is estimated by performing a linear fit to $\frac{d(g'-r')}{dlog(r)}$ or $\frac{d(B-R)}{dlog(r)}$ between $r=20$$h^{-1}_{70}$ kpc and the FWHM of the limiting PSF.  Cool-core clusters are known to populate the high $L_x$ side of the scatter in the $L_x$-$T_x$ diagram, which is where the clusters that contain blue-core BCGs are found.

\indent  The right hand panel of figure \ref{fig:LTcolour} shows the relationship between $\Delta_{L_x}$ of the host cluster and the central colour gradient of the BCG.  colour gradients measured in ($B-R$) are converted to ($g'-r'$) using the transformations of Fukugita et al. (1996).  The small red symbols represent BCGs with negative central colour gradients (red cores) and the large blue circles represent the BCGs with positive central colour gradients (blue cores).  As seen in the left hand panel of figure \ref{fig:LTcolour}, all of the BCGs with positive colour gradients (a sign of recent star-formation) are found in clusters with the highest values of X-ray excess.  Moreover, above $\Delta_{L_x}\ga0.25$ all clusters contain BCGs with positive central colour gradients.  The apparent requirement that clusters have an $L_x$ excess above the mean $L_x$-$T_x$ relation in order to host blue-core BCGs implies that the source of the cold gas required to fuel these star-forming blue-core systems is linked to the processes that give rise to the $L_x$ excess.  This result therefore points to the cooling ICM as the likely source of cold gas replenishment and not accretion from an interactions with gas rich substructure.  Furthermore, the ubiquity of star-forming cores in clusters with sufficiently high $\Delta_{L_x}$ suggest that heating in these systems (ie. AGN feedback) generally tempers the radiative cooling but does not shut it off altogether.

\subsection{Scatter in $L_x$-$T_x$ and BCG/X-ray peak offset} \label{Xraypeakoff}

\indent The results discussed in the previous sections indicate that the presence of blue-cores in BCGs is linked to the BCG's position as well as the X-ray excess of the cluster.  This is consistent with the hypothesis that cooling in the ICM is providing the cold gas necessary to fuel the recent star formation.  This star formation is one mechanism by which BCGs can acquire stellar mass, but they also grow via mergers and accretion.  Through the cluster density profile these latter two processes also depend on the BCGs cluster-centric location.  It is therefore interesting to examine to what degree the sources of scatter in the cluster $L_x$-$T_x$ relation affect the position of the BCG within the cluster or vice-versa.

\begin{figure}
   \centering
   \includegraphics[width=3.5in]{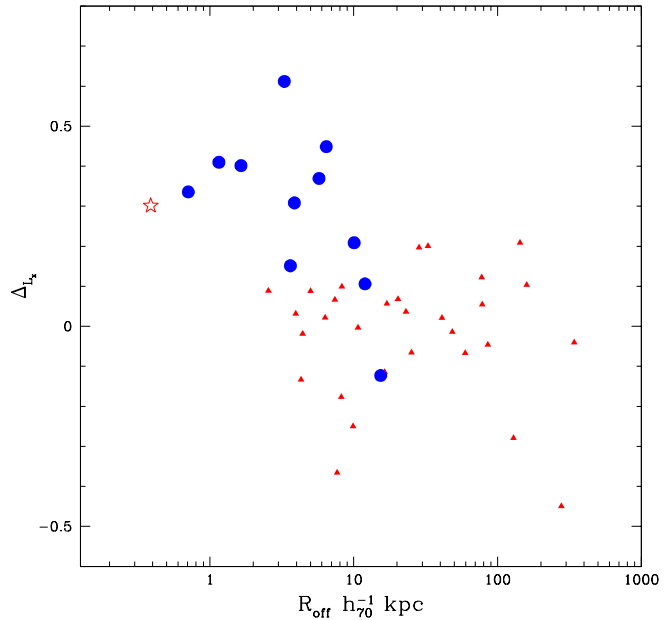} 
   \caption[$\Delta_{L_x}$ vs. BCG/X-ray offset]{$\Delta_{L_x}$ vs. BCG/X-ray offset.  Clusters with blue-core BCGs are identified with large blue circles.  Strikingly, all of the BCGs in systems with $\Delta_{L_x}\ga0.25$ are located very close to the central X-ray peak and are hosts of recent star formation.  The remainder of the systems are found at a wide range in $R_{\rmn{off}}$.  Only a single system (Abell 2261 - starred symbol) in the small offset/high $\Delta_{L_x}$ group does not contain a blue-core BCG; we find evidence in the literature, however, that suggests this system is host to dust enshrouded star formation}
   \label{fig:bcgmag_roffvsdlt}
\end{figure}

\indent  In figure \ref{fig:bcgmag_roffvsdlt} we show the dependance of cluster $\Delta_{L_x}$ on $R_{\rmn{off}}$.  The symbols in this figure are the same as those in figure \ref{fig:bcgmag_roffvsmt}.  Strikingly, all of the clusters with values of $\Delta_{L_x}\ga0.25$ have BCGs that are within $\sim 10 h^{-1}_{70}$ kpc of their cluster X-ray peaks.  Furthermore, virtually every system above this X-ray threshold host a star-forming BCG.  This is expected if the cold gas that is fueling the star formation in these systems originates through cooling of the ICM.  Abell 2261 is the only cluster with a large value of $\Delta_{L_x}$ and a small value of $R_{\rmn{off}}$ that does not contain a blue-core BCG.  As discussed in section \ref{location}, this BCG has a large dust mass $9.2 \times10^7 h_{70}^{-2} M_\odot $ (Chapman et al. 2002).  Bonamente et al. (2006) show that A2261 has a central cooling time $t_{cool} < 0.5 t_{hubble}$, as well as a central temperature gradient, indicating that it has an actively cooling core.  The existence of a system such as Abell 2261, where large amounts of cold gas remains unconverted into stars indicates that there may be additional considerations required for understanding star formation in actively cooling systems.  Perhaps there is a time delay between the build-up of fresh cold gas and the onset of star formation.  It may also be possible that an interaction with a merging substructure is required in order to trigger the star formation; or that star formation is present but hidden by the dust.  Regardless of these details, figure \ref{fig:bcgmag_roffvsdlt} demonstrates that the clusters with large positive residuals from the $L_x$-$T_x$ all have their BCGs relatively well aligned with the cluster X-ray peaks; indicating an advanced state of dynamical relaxation.

\subsection{Effect on BCG Luminosity} \label{XrayBCGlum}

\begin{figure}
   \centering
   \includegraphics[width=3.5in]{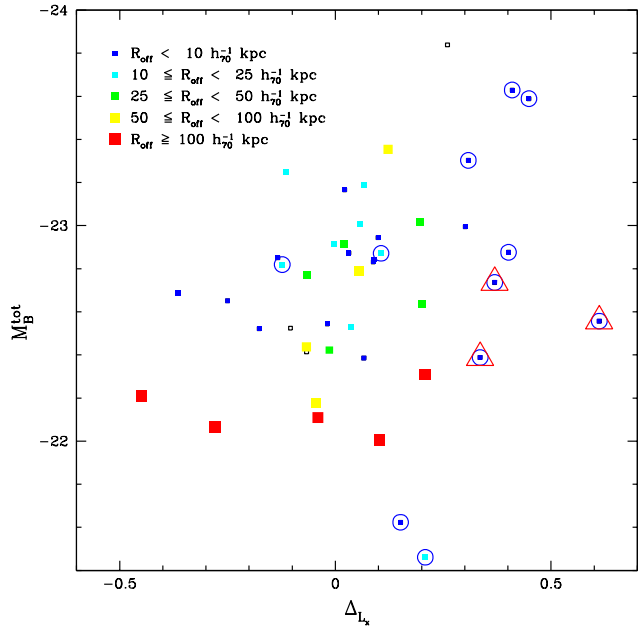} 
   \caption[BCG $M_B^{tot}$ versus cluster $\Delta_{L_x}$]{BCG total $B$-band magnitude ($M_B^{tot}$) versus $\Delta_{L_x}$ of the host cluster.  Symbols increase in size according to the BCG projected offset from the X-ray peak ($R_{\rmn{off}}$).  Blue core BCGs are encircled in blue.  Large red triangles identify the clusters with $T_x < 5$ keV.  These systems are expected to host BCGs that are less massive than those in the rest of the sample (see text).  There is a slight trend of brighter BCGs being found in clusters with an X-ray excess compared with the mean $L_x$-$Tx$ relation.  Many of the outliers from the trend have large offsets with respect to the X-ray peak.  Small open symbols represent the systems for which no X-ray data is available and so no offset is determined.}
   \label{fig:bcgmag_dlvsmt}
\end{figure}

\indent  Recent observations (Crawford et al. 1999, Wilman et al. 2006) show that some cool-core clusters are hosts of moderately high star formation rates in their centrally located BCGs (eg., Abell 1835: $SFR \sim 100 M_\odot yr^{-1}$; Abell 2204: $SFR \sim 1 M_\odot yr^{-1}$; see table \ref{tab:SFRs}).  The radiative losses in these systems are therefore not fully balanced by the heating sources; instead the energy budget is such that some of the gas is cooling enough to form stars.  Figure \ref{fig:LTcolour} illustrates that the BCGs which are forming stars are exclusively located in clusters on the high $L_x$ side of the scatter in the $L_x$-$T_x$ relation, which is the region of the diagram that is populated by cool-core clusters.  In order to further understand and quantify the importance of the cooling state of a cluster in contributing to the stellar content of BCGs, we investigate in this section how BCG luminosity varies with the X-ray excess of the host.

\indent  Figure \ref{fig:bcgmag_dlvsmt} shows BCG $M_B^{tot}$ versus $\Delta_{L_x}$.  The point size is scaled according to projected offset of the BCG centre from the peak of the X-ray distribution.  X-ray peak positions are obtained from either Chandra or XMM-Newton observations.  The bin limits in X-ray offset are 10, 25, 50 and 100 $h^{-1}_{70}$ kpc.  The points marking clusters with blue cores are surrounded with blue circles.  The clusters lacking high resolution X-ray data are plotted with open symbols.  The few clusters from the archival data sample with $T_x < 5$ keV are marked with large, open, red triangles.

\indent  Examination of figure \ref{fig:bcgmag_dlvsmt} reveals a trend of increasing BCG luminosity with increasing host cluster deviation from the mean $L_x$-$T_x$ relation.  This trend appears to be somewhat weak but if the systems with $T_x<5$keV (ie. those shown with large, open, red triangles) are ignored the trend becomes more apparent.  Furthermore, many of the outliers are BCGs which lie at large offsets from their host cluster's X-ray peak, which is expected if the $L_x$-$T_x$ deviations arise from processes in the cluster core.  Particularly striking is the absence of both faint BCGs in clusters with an $L_x$ excess and bright BCGs in clusters with an $L_x$ deficit.  A Kendall's $\tau$ test of this correlation including all data points shown in figure \ref{fig:bcgmag_dlvsmt} yields $\tau=-0.25$ with a confidence level of 98\% ($\sim 2.5 \sigma$ significance).  The trend displays a full magnitude difference for BCGs across the range in observed variation from the cluster mean $L_x$-$T_x$ relation, implying that the source(s) of scatter in the cluster $L_x$-$T_x$ relation contributes significantly to the growth of BCGs.  It is interesting that while only the clusters with the highest values of $\Delta_{L_x}$ display the observable signatures of active star formation (see figure \ref{fig:LTcolour}), the trend in BCG luminosity spans the entire range of observed X-ray excess.  This suggests that another important mechanism in addition to recent star formation is contributing to the stellar mass differentiation of BCGs.  As expected, the blue core systems inhabit mainly the upper right portion of the diagram.  The recent star formation in these systems may explain their high average luminosity with respect to the rest of the sample.  Two exceptions are the low luminosity BCGs in Abell 115N and Abell 2390, which are believed to be young BCGs that have yet to cannibalize a large portion of the galaxy population in their respective clusters (see section \ref{location} for discussion).

\indent  We examine the relationship between $M_B^{tot}$ and host cluster $L_x$ as well as $M_B^{tot}$ and host cluster $T_x$.  We find no significant evidence for a correlation between either of these two pairs of variables.

\indent The existence of the trend seen in figure \ref{fig:bcgmag_dlvsmt} ($M_B^{tot}$ vs. $\Delta_{L_x}$) suggests that whatever is responsible for producing the scatter in the $L_x$-$T_x$ relation is also assisting in the growth of stellar content of BCGs.  We have argued that cooling is the primary cause.  Alternatively, scatter may be introduced into the $L_x$-$T_x$ relation by mergers with massive substructures as well as by intrinsic variations in the shapes of the underlying dark matter halo density profiles from one halo to the next.  Recent papers by Rowley et al. (2004), Kay et al. (2006) and Poole et al. (2007) show that although merging in clusters is capable of generating some scatter in the X-ray scaling relations, this scatter is short lived and does not account for the degree of scatter seen in the observations.  In addition, work by Balogh et al. (2006) shows that the variation in dark matter density profiles also introduces some scatter but that the resulting scatter is not sufficient to reconcile the observations.  These studies therefore add support to the interpretation that the variation in BCG luminosity is directly linked to the balance between heating and cooling in the ICM.

\section{Conclusions} \label{conclusion}

\indent  We have conducted a detailed comparison of the optical properties of BCGs with their host cluster X-ray properties in 48 intermediate redshift systems using a sample that spans the full observed range in cluster morphology.  Our analysis reveals a close connection between the evolution of BCGs and that of their host clusters.  The results are discussed below.

\indent  The Kormedy Relation defined by the CCCP BCGs is found to have a steeper slope than that of regular elliptical galaxies ($a_{BCG}=3.44\pm0.13$ versus $a_E=3.02\pm0.14$).  This difference in slope may be influenced by the sample selection and the use of circular models; but it suggests that the formation processes of the CCCP BCGs differ from their regular elliptical counterparts.  Von der Linden et al. (2007) propose that the observed difference between the slope of the Fundamental Plane of BCGs and that of regular ellipticals may be a result of  an enhanced dark matter fraction due to the typical position of the BCG near or inside the cluster core.  Long-slit spectral observations of BCGs in the CCCP sample will allow us to better determine the cause of these differences and examine how baryonic feedback in clusters affects the scaling relations for galaxies. 

\indent  The majority of BCGs in the sample have shallow colour profiles that become bluer with increasing radius.  There are several BCGs (25\%) that deviate from this simple behavior; and show blue cores instead.  This colour peculiarity results in large deviations from the red sequence (up to $\sim 1.0$ mag in ($g'$-$r'$)) and affects optical cluster studies that use colour to determine cluster membership.  The radial extent of the blue-cores shows that they cannot be explained by point sources alone.  This leads us to conclude that they are regions of active star-formation.  Several of the blue-core BCGs have measured star formation rates in the literature (see table \ref{tab:SFRs}).  Dust and molecular gas has also been detected in many of these systems (Edge 2001, Edge et al. 2002, Egami et al. 2006).  These recent observations reveal that the evolution of BCGs differs from the standard picture for massive non-BCG ellipticals.  The presence of extended star forming regions in BCGs requires that the cold gas reservoir is being replenished in these systems.  Cooling flows in clusters with blue-core BCGs provide a natural mechanism to fuel this recent star formation.

\indent  One of the predictions of cool core cluster model is that the cold gas that condenses out from the ICM is deposited in the cluster centre.  We find that the star-forming systems are all located at small projected offset from the peak of the cluster X-ray emission.  Having small $R_{\rmn{off}}$ appears to be a requirement for BCGs to host a blue core; which is consistent with the cool core model prediction.

\indent  The competition between baryonic feedback effects, such as radiative cooling and heating from stellar winds/AGN, is believed to be the source of the intrinsic scatter in the cluster $L_x$-$T_x$ relation.  For a given $T_x$, radiative cooling increases $L_x$, while heating reduces $L_x$.  We find that all of the blue-core BCGs are located in clusters with large values of X-ray excess; suggesting that cooling is dominant in these systems.  The observation that all systems above $\Delta_{L_x} \ga 0.25$ contain star-forming BCGs implies that all cool core clusters are producing stars in their cores.  This is somewhat surprising and contrary to the current conventional wisdom.  Furthermore, the lack of star-forming BCGs at negative values of X-ray excess is difficult to explain with models that fail to take into account the cooling state of the host cluster.

\indent  Examination of the relationship between X-ray excess and the BCG position within the cluster reveals that all of the clusters with $\Delta_{L_x}\ga0.25$ contain BCGs (with blue cores) that are within $\sim 10$ kpc from the peak in the X-ray emission.  This suggests that the conditions responsible for the creation of cool-core clusters also lead to the formation of a centrally located BCG; both of which are required for the presence of a star-forming core.

\indent  The absolute $B$-band magnitude of BCGs is only weakly dependent upon the X-ray excess of the host cluster, as well as the BCG offset from the X-ray peak.  This is not surprising given the variety of mechanisms by which BCG acquire stellar mass (cannibalism, harassment, star formation) and the range in formation times expected from models of hierarchical merging.

\indent  These observations support a picture of linked BCG and cluster evolution.  In this scenario the clusters which contain the most dynamically relaxed BCGs are the ones that form cool cores and host central star formation.  The prototypical example of this behavior may be the cluster MS1455+22.  The optical imaging shows that it is dominated by the central cD galaxy (a blue-core BCG that is aligned with the X-ray peak) and contains very few bright satellites galaxies.  MS1455+22 appears to be in a significantly more dynamically evolved state than the majority of the clusters in the sample.  A similar conclusion is reached by Yee and Ellingson (2003) to explain its anomalous position in the velocity dispersion-richness plane.  MS1455+22 has the highest X-ray excess in the CCCP sample; indicating that it hosts a prominent cool core.  We also note that the X-ray imaging shows that the ICM in this cluster is highly relaxed.

\indent  Our results are consistent with those of Edwards et al. (2007) who find that $\sim 70\%$ of BCGs that are within 50 kpc from their host cluster X-ray peak are line emitting.  Furthermore, our results are consistent with studies that find that the recent star formation in BCGs is associated with the ICM cooling rate (Johnstone et al. 1987, Heckman et al. 1989, Cardiel et al. 1998).  However, by covering the full observed range of cluster morphologies, the CCCP sample allows us to strengthen these earlier claims and show that these phenomena are truly associated with the properties of the ICM.

\indent  Taken together, these data have several important implications.  Based on X-ray temperature and luminosity alone we can predict that if $\Delta_{L_x} \ga 0.25$ the BCG will be coincident with the cluster X-ray peak and host a blue core.  Furthermore, that blue cores are a ubiquitous feature of BCGs in cool core clusters (with sufficient X-ray excess) indicates that AGN and other heating mechanisms generally temper but do not shut off cooling altogether.  The data also show that BCG photometry alone can be used to preferentially select cool-core clusters that are prime targets for cosmological studies that focus on the cluster gas fraction ($f_{gas}$).

\indent  The results presented here are interesting and deserve further investigation.  Future work requires spatially resolved long-slit or IFU optical spectra of the CCCP BCGs.  Such data would enable one to disentangle the stellar populations in these systems and constrain the ages of the starbursts in the blue core systems.  Furthermore, because the work presented here is based solely on photometry we cannot rule out some contribution from an AGN component to the blue core.  Spatially resolved optical and near IR spectra would enable one to quantify the contribution of AGN vs. starburst as a function of position in these galaxies.

\section*{Acknowledgments}

We thank Alastair Edge, Ian McCarthy, Brian McNamara, Mike Hudson, Tod Lauer and David Spergel for helpful discussion.  We thank the anonymous referee for their useful comments.  C. B. also thanks Jon Wilis, Anudeep Kanwar and Eric Hsiao for their help with various technical issues.  This research was funded in part by the National Sciences and Engineering Research Council of Canada.

\appendix

\section{BCG Selection} \label{selection}

\indent  The task of identifying the BCG in each cluster is critical to the success of this study and for the majority of systems (ie., those containing cD galaxies) this task is straightforward.  These conspicuous galaxies have large, extended stellar envelopes and  are immediately identified as the brightest cluster member.  It is the clusters which do not contain obvious cD galaxies that pose a potential problem.  Each of the BCGs are given a selection code number from 1 to 3 corresponding to how easily they are identified, with 1 being an unambiguous selection and 3 a highly ambiguous selection.  Selection codes are listed in tables \ref{tab:MegaCamdata} and \ref{tab:CFH12Kdata}.

\subsection{Non-trivial BCG Selections}

\indent  There are clusters for which two or even three galaxies are potential BCG candidates.  In order to select one of these candidates over another the additional qualification must be made that the galaxy of interest is the candidate which is most likely to have been influenced by the cluster's gravitational potential throughout its evolutionary history.  This additional qualification allows us to take advantage of the outstanding high-resolution X-ray observations available in the Chandra and XMM-Newton archives.  Using these data we select the candidate that is located closest to the X-ray centroid of its host cluster as the BCG.  For a relaxed cluster, the X-ray centroid is expected to trace the deepest point in the cluster potential.  In the following discussion we describe the systems with multiple BCG candidates and outline the reasoning used in the final BCG selection.

\vskip 6pt
\noindent $\bullet$  Abell 370: The X-ray emission from this cluster is peanut shaped with two lobes connected by a bridge.  Each lobe has an associated BCG candidate.  This system is likely a pair of sub-clusters in the early stages of a merger and thus we treat both candidates as individual BCGs with the north (N) and south (S) BCGs labeled 'a' and 'b' respectively.

\vskip 6pt
\noindent $\bullet$  Abell 520: This cluster has a highly disturbed X-ray morphology indicative of a major merger.  There are two BCG candidates lying on the outskirts of the cluster X-ray contours.  The NE candidate is chosen as the BCG over the SW candidate because it lies closer to the centroid of the overall X-ray distribution.

\vskip 6pt
\noindent $\bullet$  Abell 959 contains several BCG candidates.  There are two galaxies in particular that lie at small projected distances from the X-ray peak.  Of these two we choose the N candidate as the BCG because it lies slightly closer to the X-ray centroid.  We note that this selection is somewhat uncertain because the X-ray image that is used to arrive at this selection is from Rosat.

\vskip 6pt
\noindent $\bullet$ Abell 1234: There are two potential BCG candidates in this clusters.  They are identified as the NW candidate and the SE candidate.  There are no X-ray data for this cluster in either the XMM-Newton or Chandra archives.  We select the NW candidate as the BCG because it is more centrally located within the cluster and it appears to have a more extended stellar envelope in comparison with other candidates.

\vskip 6pt
\noindent $\bullet$ Abell 1246: This cluster contains two potential BCG candidates.  These object are overlapping on the sky and distinguished as the N candidate and S candidate.  The S candidate has a higher peak surface brightness.  The N candidate however has a position angle that is more aligned with the overall extended profile of these two components.  The extended profile also seems to be more centred on the N candidate.  We choose the N candidate as the BCG.

\vskip 6pt
\noindent $\bullet$ Abell 1914: This cluster has two BCG candidates with one in the NE and the other in the SW.  The SW candidate has a brighter peak surface brightness than the NE candidate but it lies further from the X-ray centroid by a significant margin.  We choose the NE candidate as the BCG.

\vskip 6pt
\noindent $\bullet$  Abell 2111: This cluster has a pair of BCG candidates.  We select the N candidate as the BCG over the S candidate because it is closer to the X-ray peak and it has a much more extended low surface brightness envelope.

\vskip 6pt
\noindent $\bullet$  CL0910+41: This cluster has a very bright BCG that sits near the centre of the optical light distribution (SDSSJ091345.5+405628).  This galaxy is unlike many other BCGs in that it is compact, extremely blue in colour and contains very little evidence of an extended stellar envelope.  These differences raise the concern that this galaxy may be a foreground source that projects onto the cluster.  Goto (2005) identifies this galaxy as a Hyper Luminous InfraRed Galaxy (HLIRG) with an infrared luminosity $L_{ir}>10^{13}L_\odot$ at $z=0.442$ which makes this galaxy a confirmed cluster member and validates our BCG selection in this case.  The extraordinary properties of this galaxy as a BCG make it particularly interesting.

\vskip 6pt
\noindent $\bullet$ MS0906-11: The galaxy chosen as the BCG in this cluster was selected based on its small offset from the central X-ray peak as well as the flat shape of its surface brightness profile compared to other potential BCG candidates in this cluster.

\vskip 6pt
\noindent $\bullet$ MS1621+26: The X-ray image shows a disturbed morphology containing an elongated central peak with a plume of emission extending from the central peak to the north.  Of the two BCG candidates (N and S) we select the southernmost candidate because it lies closer in projection to the central peak while the northernmost candidate appears to be more associated with the plume of emission.  It is also interesting to note that the selected BCG has a double peaked core visible in the R image.  This system appears to be in an intermediate stage of a merger between two sub-clusters.

\subsection{Clusters with strong BCG Ambiguities}

\indent After selecting BCGs in clusters for which the selection is not immediately obvious there still remains a group of clusters in which multiple BCG candidates are similar enough according to our selection criteria that it is not possible to reliably label a single galaxy as the dominant BCG.  In the following we briefly describe the systems for which this is the case.

\vskip 6pt
\noindent $\bullet$  Abell 851 contains several candidate BCGs.  We narrow the selection to a closely spaced group of three bright galaxies that lie nearest the centre of X-ray emission in an XMM-Newton 24 ks exposure.  Of these three remaining candidates we select the southernmost galaxy as the BCG because it appears to have the largest extended stellar envelope.

\vskip 6pt $\bullet$ CL0024+16:  This cluster has a complex core with four bright, red galaxies spaced with their surface brightness profiles significantly overlapping such that they appear to share a common diffuse halo.  It is likely that these galaxies are in the processes of merging to form a cD galaxy.  A 40ks Chandra image of this field shows a mostly relaxed X-ray surface brightness profile which is elongated in the direction of the four bright, red, central galaxies.  The easternmost of the four red galaxies is coincident with the X-ray peak however the active dynamical state of the system makes the selection of the BCG based on the X-ray distribution less reliable because of possible separation between the gas and the red galaxies.  The combined difficulty in BCG selection and modeling of the overlapping red galaxy surface brightness profiles cause this cluster to be flagged and removed from some of the analyses.

\vskip 6pt $\bullet$ MS0016+16: This cluster contains three potential BCG candidates which in projection are seen to lie along a line from the northeast to southwest.  The two most northeast candidates lie closest to the central X-ray peak of the cluster but are both at the same projected clustercentric distance.  It is likely that these three systems will eventually merge to form the BCG but in the current dynamical state and viewing angle its is not possible to reliably select a BCG from this cluster.

\vskip 6pt $\bullet$ MS1231+15:  There are two obvious BCG candidates in this cluster which appear to be in the intermediate stages of merging together.  They share a common halo of diffuse emission that is elongated along the axis connecting the two galaxies.  There is no X-ray image of this cluster in either the Chandra or XMM archives and thus it is impossible to select a BCG based on the shape of the X-ray distribution of this cluster.  These two candidates are labeled MS1231a and MS1231b (north and south respectively) but are left out of some of the analyses because of the large ambiguity in BCG selection and the lack of X-ray imaging.

\section{Analysis} \label{analysis}

\indent  We remove the image background using a standard 2-d polynomial model.  While a more complex background subtraction method would improve the depth of our surface brightness measurement, such an improvement does not significantly affect the values of the BCG fit parameters.

\indent For each cluster image we generate a mask which is used to reject pixels from inclusion in the surface brightness measurement of the BCG.  The mask is based on a combination of 3 layers and is designed to mask everything but the pixels that are unambiguously identified with BCG or background light.

\indent  The first layer of the mask is derived from a SExtractor segmentation image.  The initial choice of extraction parameters are based on those used in Simard et al. (2002).  The segmentation image produced is then modified by extending each masked region into the background by 3 pixels on all sides in order to reduce the contamination from the unmasked portion of the detected sources lying below the detection threshold.

\indent  The second layer of masking is generated by hand and is used to exclude regions of the image that are obviously contaminated by sources and defects that are not detected by automated source extraction routines such as SExtractor (Bertin \& Arnouts 1996).  Most of the image stack is constructed from at least 4 separate exposures however, due to small gaps between adjacent chips on the mosaic detector there are sections of the stack which contain only 3 or fewer exposures depending on the dither pattern used.  With fewer than 4 exposures it is difficult to determine a robust median value when stacking and as a consequence there are more cosmic rays present as well as abrupt changes in the signal-to-noise properties across these regions.  To avoid complications in surface brightness fitting these regions are also masked in this layer.

\indent  Finally, the central regions around the BCG are masked by hand out to a radius of 45" for each target BCG.  Each image is carefully examined and all visible sources are masked out to the radius at which they become indistinguishable from the background level.  The contrast is adjusted slightly and the process is iterated to fainter and fainter surface brightness.  The resulting mask portion is used as the third mask layer and replaces the isophotal layer ($2^{nd}$ layer) within 45" of the BCG centroid.  This third layer of masking is required to ensure that galaxies in the highly crowded cluster cores do not contaminate the fit.  Isophotal masking cannot account for the shapes of the core galaxies and so it is more effective to mask them by hand as the eye can trace the falling surface brightness profiles out to large distances quite accurately.  We have also experimented with modeling and subtracting the core galaxies prior to BCG surface brightness measurement.  As Patel et al. (2006), however, we find that this technique does not improve the surface brightness measurements over those obtained by simple masking.

 \begin{figure*}
    \centering
\begin{tabular}{c c}
    \includegraphics[width=3.4in]{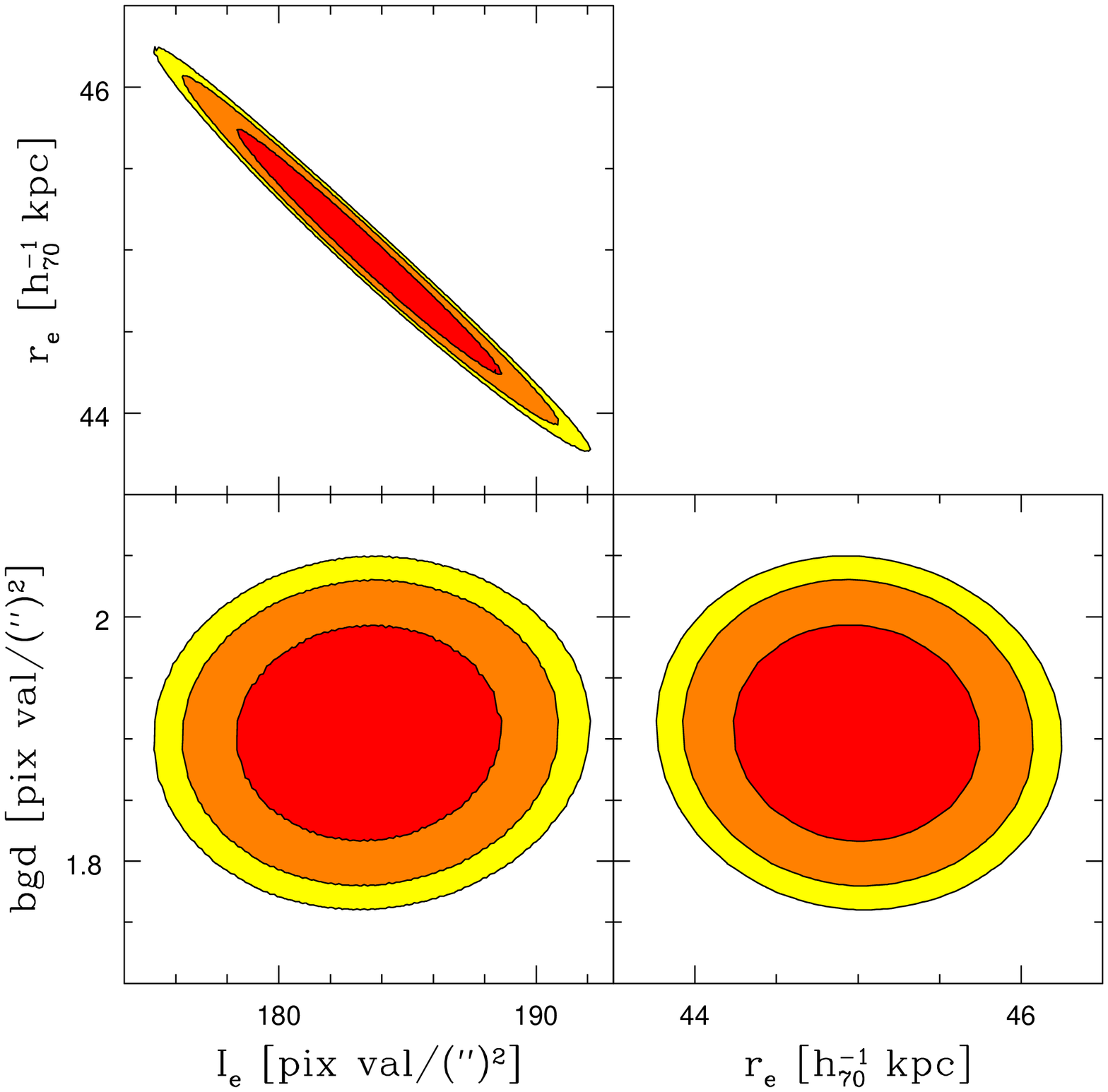} & \includegraphics[width=3.4in]{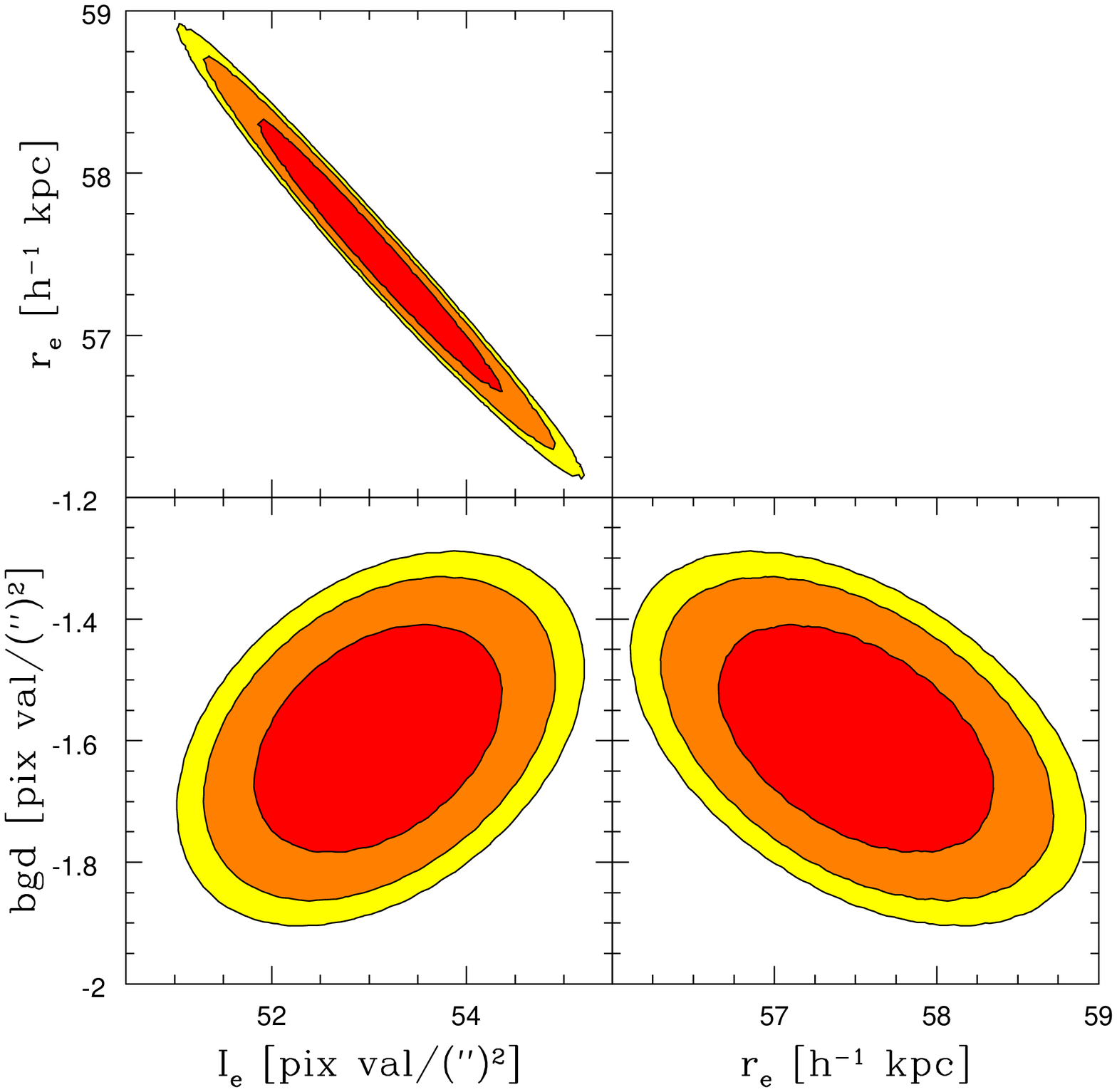}
\end{tabular}
    \caption[$\chi^2$ contours of surface brightness fit parameters]{Marginalized $\chi^2$ contours for surface brightness fit parameters  on $r'$ (left) and $g'$ (right) images of Abell 2259.  The contours are shaded according to confidence level: $1\sigma$ in red, $2\sigma$ in orange and $3\sigma$ in yellow.  The most notable degeneracy is between $r_e$ and $I_e$.  We note that the dependence of the fit parameters on wavelength is due to the presence of a colour gradient in this system.}
    \label{fig:chi2surf2259}
 \end{figure*}

\subsection{BCG surface brightness modeling} \label{sbfit}

\indent  For each BCG surface brightness profile a PSF-convolved, circularly-symmetric, single-component $r^{1/4}$ deVaucouleurs model is fit with the following functional form:

\begin{equation}
I(r)=I_e \exp{(-7.67((r/r_e)^{1/4}-1))}+I_{bgd}
\label{eqn:dev}
\end{equation}

\noindent with $I_e$ and $r_e$ as the effective intensity and effective radius respectively.  The term $I_{bgd}$ is a constant background intensity term which is also included as a fit parameter.  To determine the PSF we measure the surface brightness profile of an unsaturated star close to the BCG.  The total integrated distribution is then normalized to unity in order to conserve total flux when applying the PSF as a convolution response function in BCG modeling.  We fit a moffat profile (Moffat 1969) to the PSF star with the following form:

\begin{equation}
I(r)=I_0[1+(r/\alpha)^2]^{-\beta}+I_{bgd}
\label{eqn:moffat}
\end{equation}

\noindent where $I_{bgd}$ is a term added to take into account the local background level and $\beta$ is left as a free parameter.

\indent Estimates for the error in surface brightness for each radial bin are obtained by bootstrapping the pixel values in that bin.  BCG source pixels are randomly selected in each radial bin, allowing a given pixel to be selected multiple times.  This has the advantage that we do not need to make any apriori assumption of the functional form of the noise (ie. poisson noise), or to know the precise values of exposure times, gain or read-noise.  Bootstrapping not only accounts for the random noise in the data, it also incorporates some of the contribution from non-uniform, systematic variations (bright stars, variability in galactic dust absorption, crowding effects of nearby galaxy/group light, etc.) across the cluster image.  We note, however, that bootstrapping estimates misrepresent the true error in radial bins that have only a few pixels available to randomly sample.
 
\indent  The BCG fitting procedure is restricted to data points at radii larger than twice the FWHM of the derived moffat profile PSF ($r_{fit} > 2 r_{fwhm}$).  Ignoring the central bins helps to ensure that any small error in the PSF determination will not seriously affect the BCG fit.  An upper radial cutoff to the fitting region is also applied.  The fit is initially restricted to those data points that lie at radii less than $500 h^{-1}_{70} kpc$ from the BCG centre.  This radius is chosen somewhat arbitrarily but it reflects an effort to isolate the BCG component within the overall cluster diffuse light profile.  The PSF-convolved deVaucouleurs + background model is then fit to the data lying within these two limits as described above.  A second iteration of the fit is then performed with the maximum fit radius set to 1.5$r_{e,initial}$ and an extra term is added to $\chi^2$.  The second fit iteration uses a $\chi^2$ calculated as follows:

\begin{equation}
\chi^2_{new}=\chi^2_{initial}+\bigg(\frac{I_{bgd,new}-I_{bgd,initial}}{\sigma_{bgd,initial}}\bigg)^2
\end{equation}

\noindent where the subscripts $initial$ and $new$ refer to the first and second fit iteration parameters respectively.  The extra background term in $\chi^2_{new}$ acts as a penalty to any major change in the background parameter and ensures that the background is primarily determined/constrained by the larger radius fit range.  The reduction in the fit range of the second fit iteration helps to further isolate the BCG signal with an upper radial cutoff that is motivated by the shape of the galaxy itself.

\indent  Examples of the marginalized $\chi^2$ contours are shown in figure \ref{fig:chi2surf2259} for the fits to the Abell 2259 BCG surface brightness profile.  It is immediately seen that $I_e$ and $r_e$ are the most degenerate parameter pair, highlighting a common trait of all of BCG surface brightness fits in the CCCP sample.  By leaving the background as a free parameter in the fitting routine, even though the image has been background subtracted, this technique shows how the remaining structural properties of the BCG are affected by a possible error in background.  By inspection of the $\chi^2$ contours shown in figures \ref{fig:chi2surf2259} it can be seen that a 3$\sigma$ error in the background can affect the derived $r'$ effective radius $r_e$ at the 3-4 \% level for Abell 2259.  The marginalized $\chi^2$ surface diagrams are also useful for analyzing the stability of the fitting routine for a given BCG.  In summary, these data indicate that our BCG fitting procedure determines the structural parameters with high precession (within a few \%) and that the effects of error in the background subtraction are negligible.

\section{Optical Photometry} \label{photo}

%SURFACE BRIGHTNESS PROFILES

\begin{figure*}
   \centering
   \includegraphics[width=7.0in, height=8.4in]{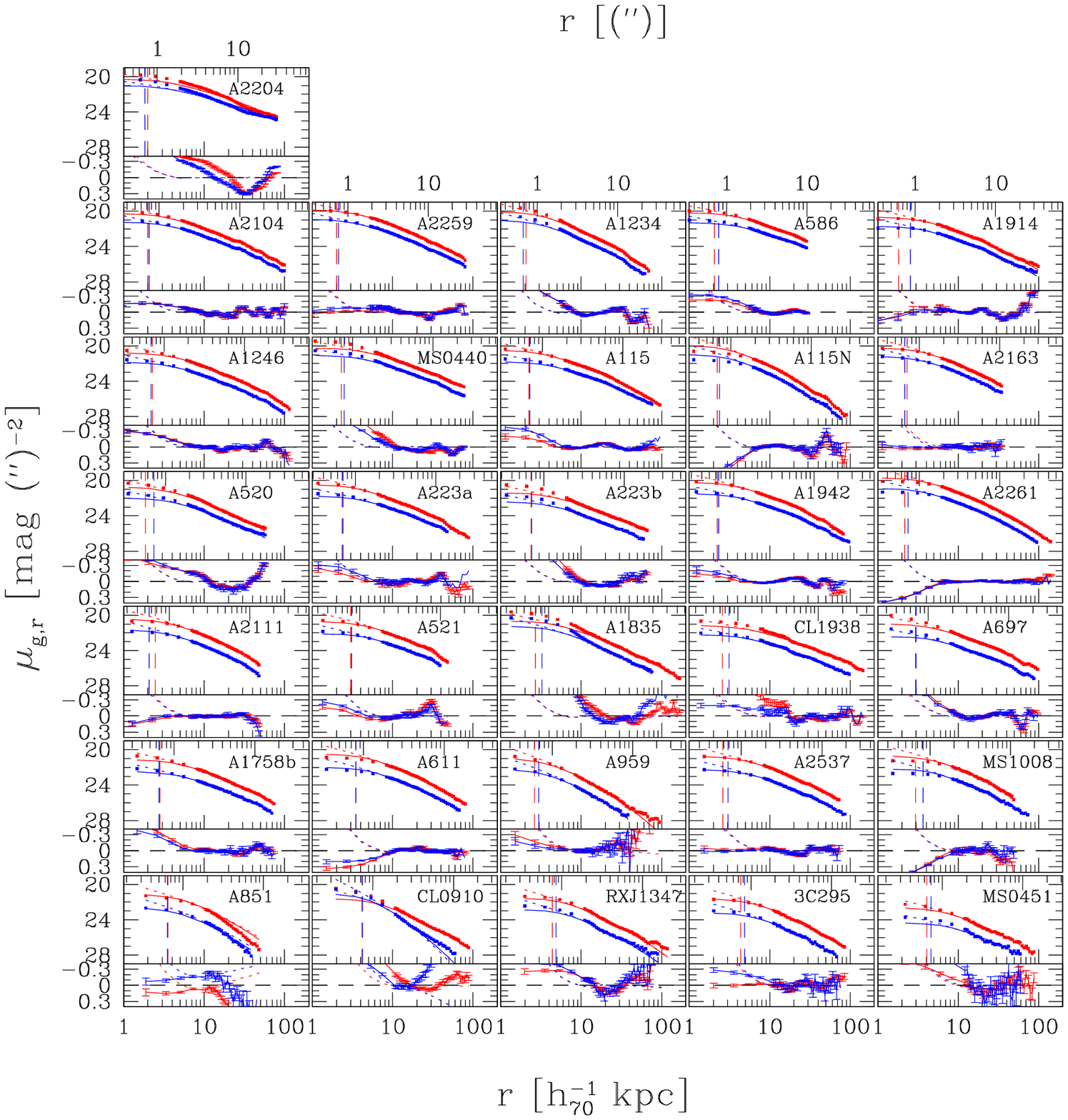}
   \caption[Surface brightnesses in $g'$ (blue) and $r'$ (red) for BCGs]{Surface brightnesses in $g'$ (blue) and $r'$ (red) for the MegaCam BCGs in the CCCP sample.  Points correspond to measured data while solid and dotted lines correspond to fitted $r^{1/4}$ models with and without PSF-convolution respectively.  Vertical dashed lines denote the $r_{fwhm}$ seeing limit of the derived moffat PSF.  Residuals are shown along the bottom with dotted lines showing the difference between fits with and without PSF-convolution and solid lines with error bars showing the difference between PSF-convolved models and the data.  The top axis shows the corresponding radii in arcseconds}
   \label{fig:sb31_mega}
\end{figure*}

\begin{figure*}
   \centering
   \includegraphics[width=7.0in, height=8.4in]{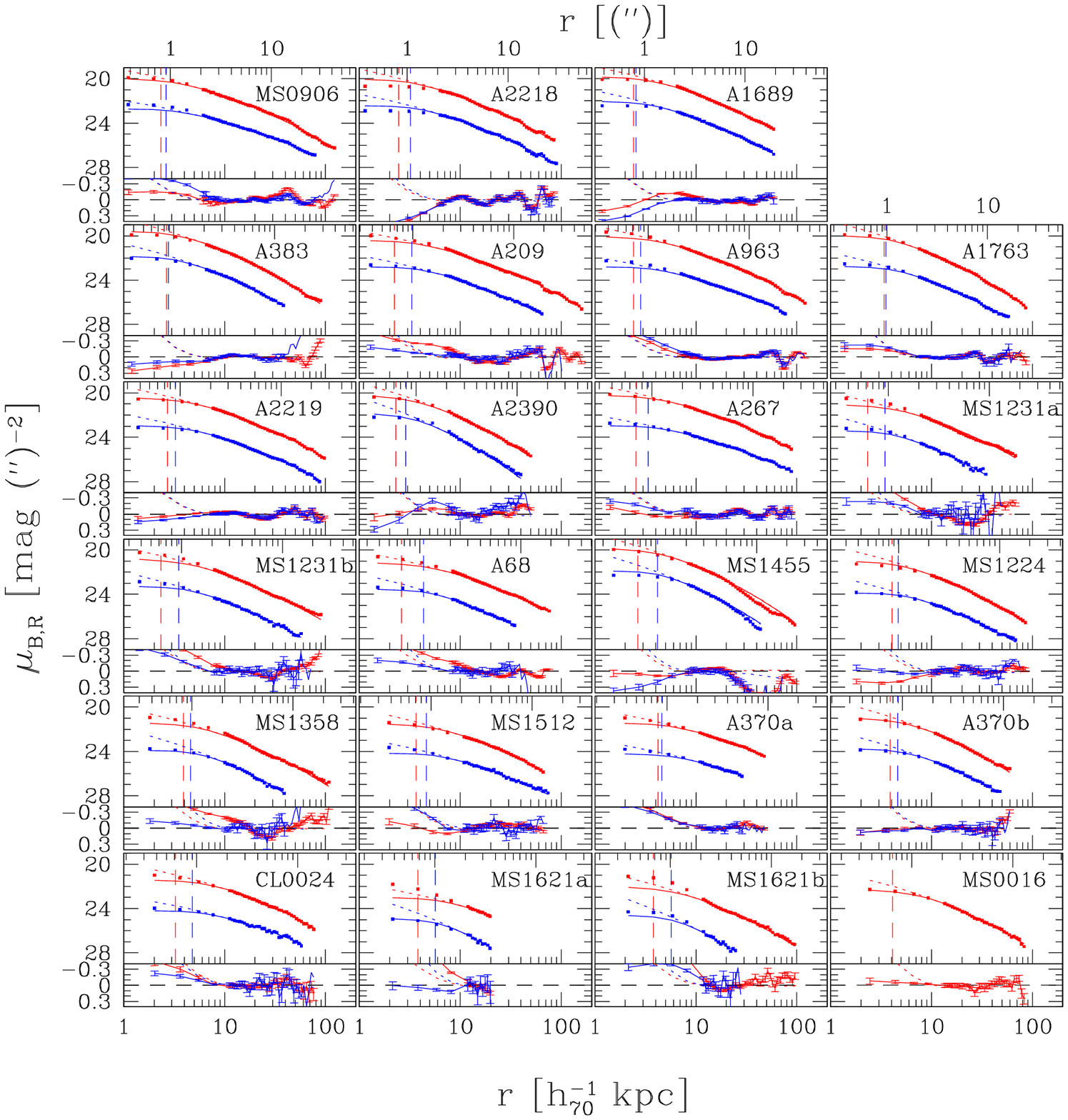}
   \caption[Surface brightnesses in $B$ (blue) and $R$ (red) for BCGs]{Surface brightnesses in $B$ (blue) and $R$ (red) for CFH12K BCGs in the CCCP sample.  Points correspond to measured data while solid and dotted lines correspond to fitted $r^{1/4}$ models with and without PSF-convolution respectively.  Vertical dashed lines denote the $r_{fwhm}$ seeing limit of the derived moffat PSF.  Residuals are shown along the bottom with dotted lines showing the difference between fits with and without PSF-convolution and solid lines with error bars showing the difference between PSF-convolved models and the data.  The top axis shows the corresponding radii in arcseconds}
   \label{fig:sb23_12k}
\end{figure*}

%colour PROFILES

\begin{figure*}
   \centering
   \includegraphics[width=7.0in, height=8.4in]{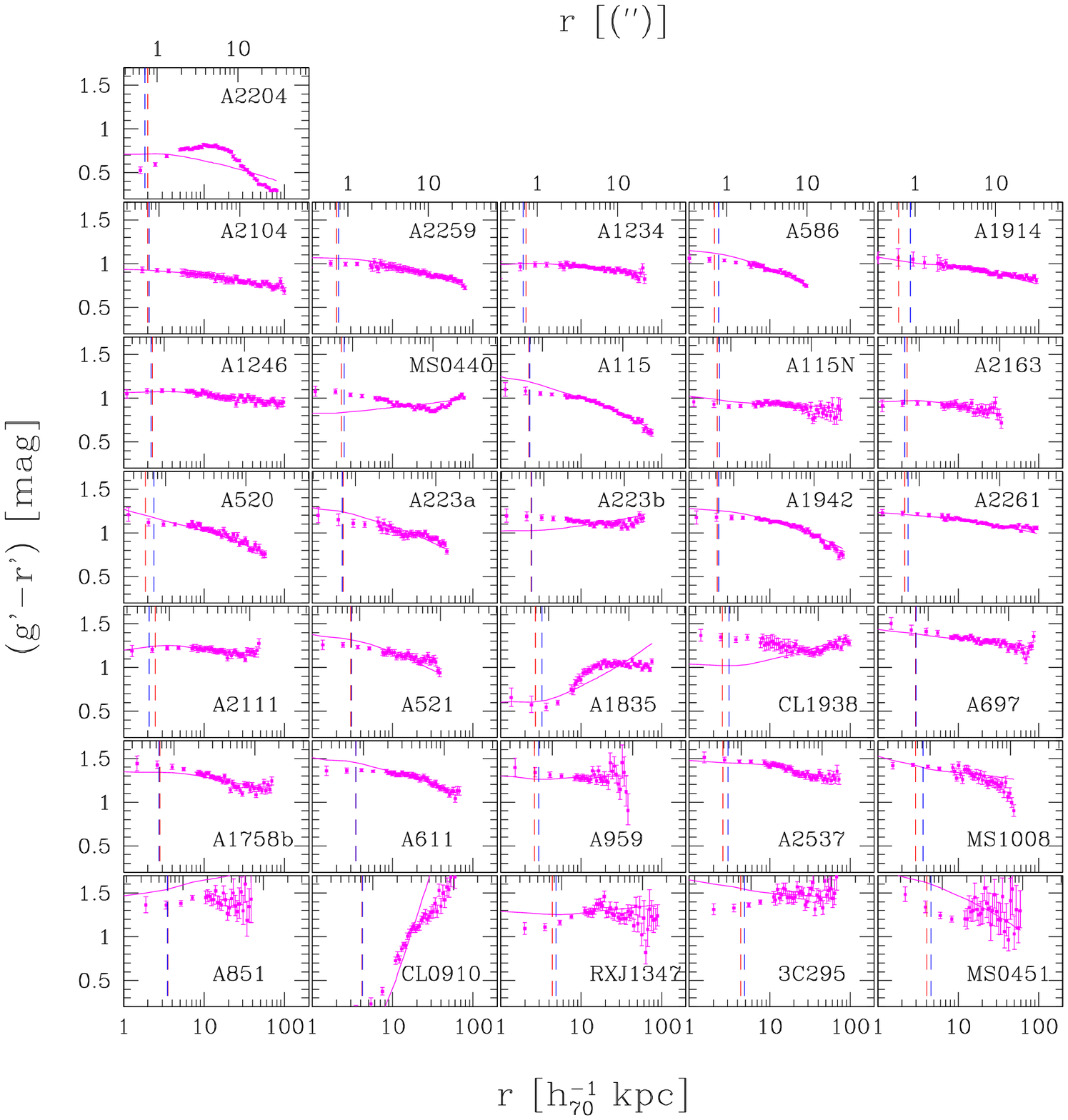} 
   \caption[colour profiles in ($g'$-$r'$) for BCGs]{colour profiles in ($g'$-$r'$) for BCGs observed with MegaCam in the CCCP sample.  Points with error bars indicate measurements.  Solid lines show the difference between best fit $g'$ and $r'$ $r^{1/4}$ models.  Vertical dashed lines show the seeing limits $r_{fwhm}$ for the $g'$ data (blue) and $r'$ data (red).  The top axis shows the corresponding radii in arcseconds}
   \label{fig:c31_mega}
\end{figure*}

\begin{figure*}
   \centering
   \includegraphics[width=7.0in, height=8.4in]{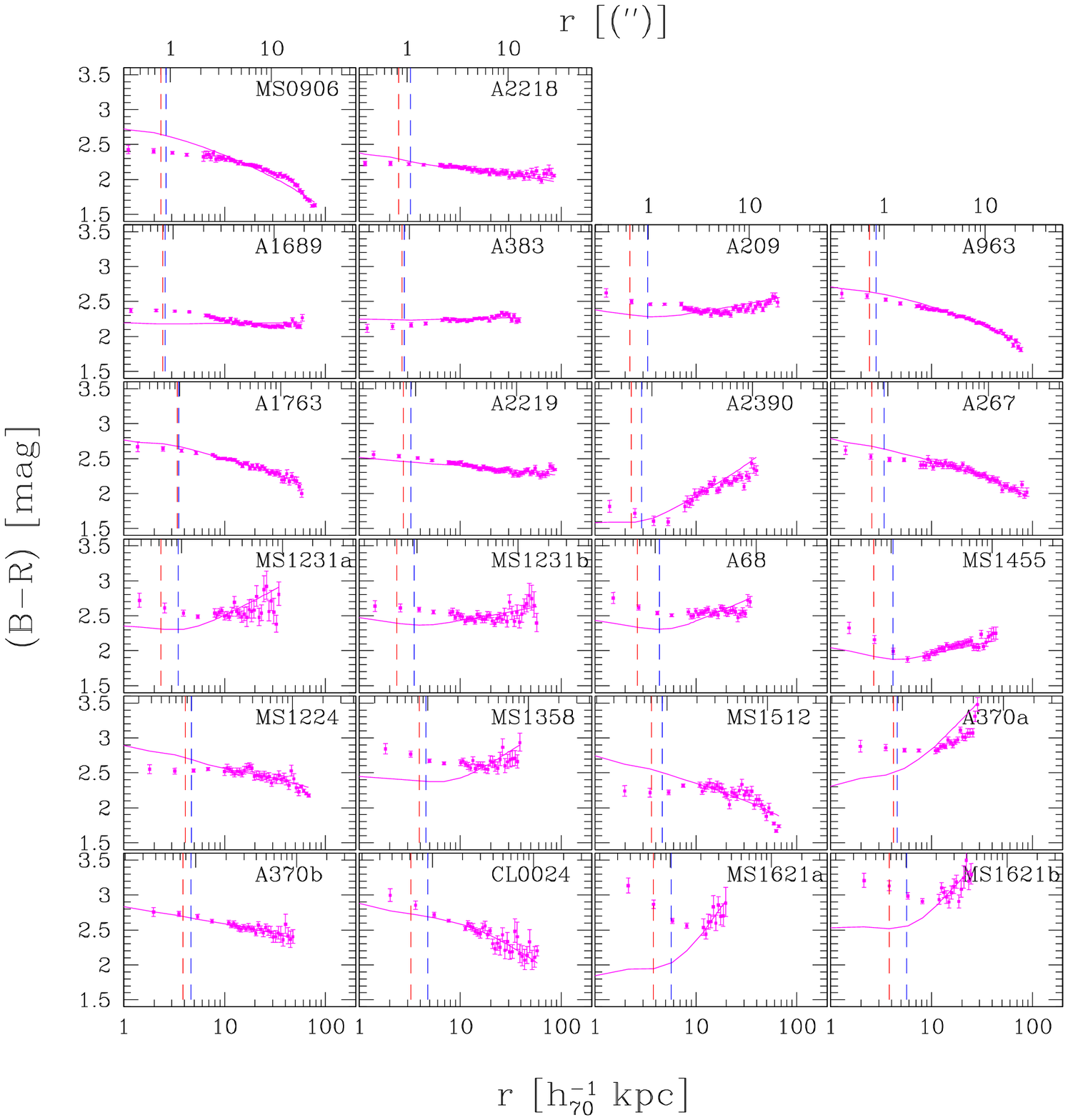} 
   \caption[colour profiles in ($B$-$R$) for BCGs]{colour profiles in ($B$-$R$) for BCGs observed with CFH12K in the CCCP sample.  Points with error bars indicate measurements.  Solid lines show the difference between best fit $B$ and $R$ $r^{1/4}$ models.  Vertical dashed lines show the seeing limits $r_{fwhm}$ for the $B$ data (blue) and $R$ data (red).  The top axis shows the corresponding radii in arcseconds}
   \label{fig:c22_12k}
\end{figure*}

\indent  Figure \ref{fig:sb31_mega} shows the surface brightness profiles in the $g'$ and $r'$ bands for all BCGs that were observed with MegaCam while figure \ref{fig:sb23_12k} shows the surface brightness profiles of all BCGs that were observed with CFH12K.  The measured data are shown by the points, solid lines correspond to best fit PSF-convolved $r^{1/4}$ models as described in the above section and dotted lines show similar fits without PSF convolution.  Surface brightness data is colour coded by wavelength band with $r'$ and $R$ data shown in red and $g'$ and $B$ data shown in blue.  In all cases the BCGs are brighter in the red filter than in the blue filter with a typical offset on the order of one magnitude at all radii.  Vertical dashed lines show the moffat PSF fwhm radius ($r_{fwhm}$).  Residuals are shown along the bottom of each plot.  In each sub-panel the dotted line corresponds to the difference between the PSF-convolved and non-PSF-convolved fits, and the solid lines with error bars show the difference between the PSF-convolved models and the measured data.  Note that the $B$-band data for MS0016 has been omitted because it was taken under non-photometric conditions.  See section \ref{sbprodiscuss} for discussion.

\indent  Figure \ref{fig:c31_mega} shows the colour profiles in ($g'$-$r'$) for the MegaCam BCGs, while figure \ref{fig:c22_12k} shows the colour profiles in ($B$-$R$) for the CFH12K BCGs.  As before, the vertical dashed lines denote the $r_{fwhm}$ of the PSFs in $r'$ or $R$ (red) and $g'$ or $B$ (blue).  Points with error bars indicate the measured data and solid lines show the difference between the red and blue $r^{1/4}$ model fits to the surface brightness distribution.  Note that even though this is not a direct fit to the colour profile the difference between the individual fits to the data frames match the colour data well in most cases.  See section \ref{colourprodiscuss} for discussion.

\section{Quantifying Systematics}

\begin{figure*}
   \centering
   \includegraphics[width=7in]{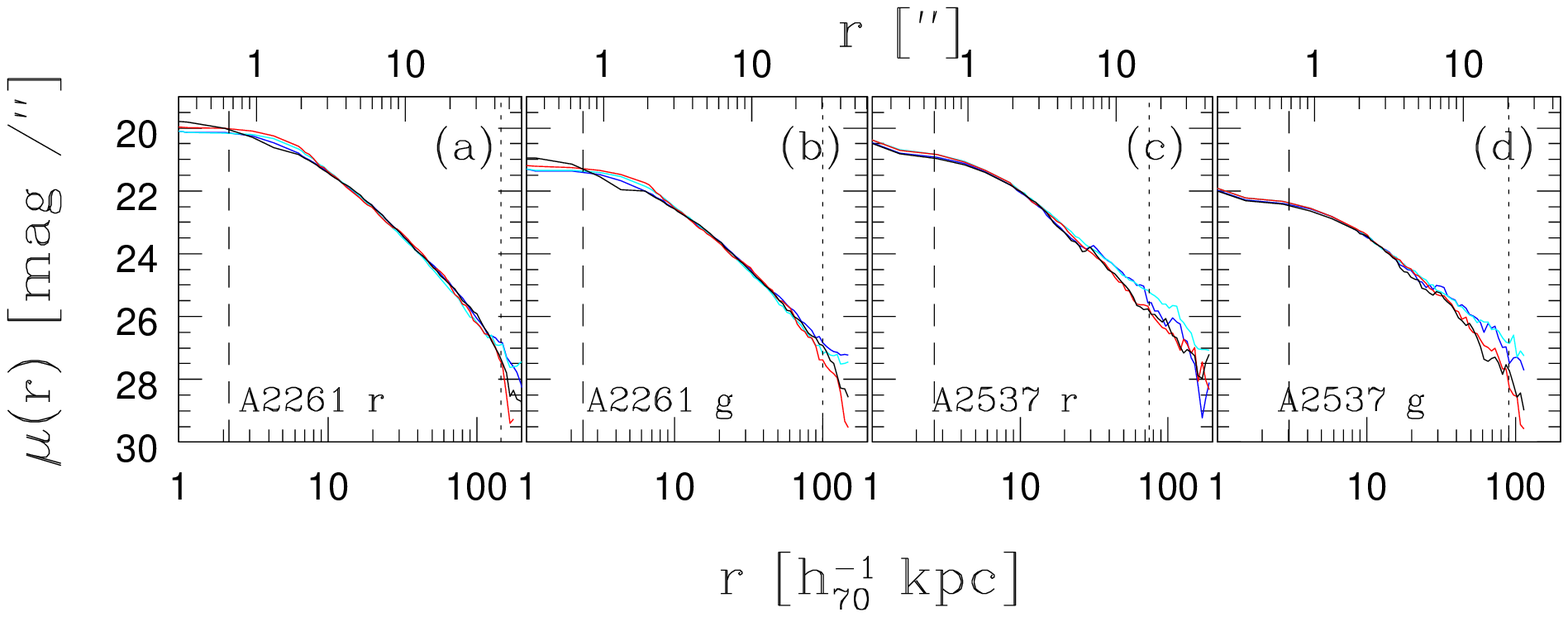} 
   \caption[Masking effects tests]{Testing for systematic error introduced by mask effects in Abell 2261 and A2537.  The BCGs in these clusters are divided along their major and minor axes into four quadrants.  Each of the different coloured curves in these diagrams represent the surface brightness measured in a single quadrant.  These diagrams are used to determine the radial range in which the flux can be reliably assigned to the BCG (see discussion).  Large differences in the profiles from the four quadrants indicate asymmetric sources of contaminating flux.  The panels (a) and (b) show respectively the $r'$ and $g'$ surface brightness profiles for the BCG in Abell 2261.  The panels (c) and (d) show the same for the BCG in Abell 2537.  The dashed line shows the FWHM of the moffat profile seeing disk.  The dotted line shows the radial limit to which the data are reliable.  Beyond this limit the data are systematically biased by contamination from mask effects.  Identical tests are performed on the $B$ and $R$ profiles for the clusters observed with CFH12k.}
   \label{fig:mask_incomp2.eps}
\end{figure*}

\indent The measurement of surface brightness profiles of BCGs requires a detailed understanding of the various sources of systematic effects that may contaminate the signal.  Previous BCG and ICL studies have employed a wide variety of tests that aim to quantify these sources of contamination (Gonzalez et al. 2005, Krick et al. 2005, Patel et al. 2006).  The following sections describe the tests that are used in this paper to investigate the contributions of systematics effects and the steps taken to minimize them.

\subsection{Mask Effects}

\indent The measurement of BCG surface brightness is complicated by the effects of crowding.  The shallow surface brightness profiles of other luminous galaxies in the field often overlap with that of the BCG.  The method used to create the mask therefore systematically affects the measured profile.  The most common mask generation method used in the literature rejects pixels around non-BCG galaxies that lie inside a particular isophote $\mu_{iso}$ (Lin and Mohr 2004, Feldmeier et al. 2004, Zibetti et al. 2005, Krick et al. 2005).  The advantage of this technique is that the methodology is clearly defined and reproducible, the disadvantage however, is that it does not take into account the profile shapes.  The mask generation method used here is not limited to a particular isophotal cutoff and faint shallow wings of BCG neighbors are masked extensively.  In this section we discuss the tests used to evaluate the influence of mask effects on surface brightness.

\indent The test described here is designed to look for signal contamination by comparing the profiles in different angular sections of the BCG.  The image is divided along the BCG major and minor axes into four quadrants.  Under the assumption of zero contamination, the profiles measured in each quadrant should be identical.  The differences between the profiles in different quadrants is a measure of the asymmetric component of the contamination signal.  There are other causes of variation in the quadrant profiles such as asymmetric masking and position angle twists.  For instance, if an individual quadrant is heavily masked near the minor axis then the measured profile in that quadrant is more representative of the surface brightness along the major axis and is brighter than the ideal unmasked, zero contamination case.  In this way, slight variation in the masking from one quadrant to another introduces scatter among the four quadrant profiles.  This test is applied to all BCGs and the presence of large scatter in the quadrant profiles is used to determine the radial limit to which the BCG surface brightness data are reliable in each of the images ($r_{dat,max}$).

\indent The plots shown in figure \ref{fig:mask_incomp2.eps} show the result of the masking incompleteness tests for the BCGs in the clusters Abell 2261 and Abell 2537.  The profiles in the four quadrants are shown as different coloured curves.  Dotted lines indicate the radial limit ($r_{dat,max}$) to which the the BCG profile is minimally affected by asymmetric mask incompleteness.  The profile deviations seen in the $r'$ image of Abell 2537 just inside $r_{dat,max}$ are due to a position angle twist of the BCG.  This effect is symmetric in two of the four quadrants and anti-symmetric in the other two.

\subsection{Large Scale Background and Flatfield Error \label{bgdvar}}

\indent Another major source of systematic error in BCG/ICL studies comes about from large scale flatfielding error and large scale variations in the background.  Feldmeier et al. (2002) and Patel et al. (2006) present methods that can be used in order to quantify some of the properties of background variation on a variety of scales.  Similar methods are used here in order to quantify the image background variability.  Each image is portioned into a rectangular grid.  The background statistics are determined within each grid square using the same mask that is used to measure BCG surface brightness.  This process is repeated with variety of grid sizes in order to examine background variation on different scales.  The results of this procedure reveal the presence of structure in the background as a function of distance from the BCG centre and size of the grid.

\indent  To quantify the effects of large-scale structure we examine the statistics of 10 independent surface brightness profiles that are centreed on various parts of the image background.  From the analysis of several cluster images, we conclude that the variance in the background signal begins to dominate the surface brightness measurements at a typical limit of $\mu=28.5$ $mag/(")^2$ but that brighter than this limit, large-scale error does not significantly contribute to the BCG signal.

\indent  We test for the possibility that some of the BCG/ICL might be removed in the background subtraction process due to its large extent.  Examining a range of mesh sizes used for background determination, we conclude that this is not an important effect for the results presented here (5\% difference in $r_e$ in the most extreme case).

\bsp

\label{lastpage}

\end{document}